\documentclass[11pt,english,british,fleqn]{article}
\usepackage[T1]{fontenc}
\usepackage[latin1]{inputenc}
\usepackage{babel}

\usepackage{amssymb}
\usepackage[unicode=true, pdfusetitle,
 bookmarks=true,bookmarksnumbered=false,bookmarksopen=false,
 breaklinks=false,pdfborder={0 0 1},backref=false,colorlinks=false]
 {hyperref}

\makeatletter



\makeatother

\begin{document}

\title{The conflict between realism and the scalar potential in electrodynamics.}

\author{D. F. Roscoe \\
 School of Mathematics and Statistics \\
 University of Sheffield \\
 Sheffield S3 7RH, UK}
\maketitle
\begin{abstract}
Within no inertial frame can \emph{stationary charge} exist. All charge,
wherever it exists, experiences perpetual interaction with charge
elsewhere and so can only exist as non-trivial current. It follows
that the notion of the electrostatic scalar potential is a pragmatic
idealization - it is not fundamental and is useful simply because,
in many practical circumstances, the time scales of interest are sufficiently
small that charges can be considered stationary in appropriately chosen
inertial frames. In this paper, we take the view that the subsuming
of the electrostatic scalar potential within classical electrodynamics
raises issues of fundamental principle and consider the question of
whether it is possible to have a relativistic electrodynamics \emph{sans}
the electrostatic scalar potential - that is, expressed purely in
terms of some form of \emph{relativistic} magnetic potential. Surprisingly,
we find that such a thing can be realized in a remarkably simple way
- and which is cost-free for ordinary applications of electrodynamics.
But the overall formalism has fundamental ramifications for our deeper
understanding of electrodynamics - for example, a new relativistic
symmetry arises which, naturally interpreted, implies that the Lorentz
force must be considered as a local-action contact force in which
energy-momentum is locally conserved through the medium of an uncharged
massive vector particle.
\end{abstract}

\section{Introduction\label{sec.Intro}}

Within no inertial frame can \emph{stationary charge} exist. All charge,
wherever it exists, experiences perpetual interaction with charge
elsewhere and so can only exist as non-trivial current. It follows
that the notion of the electrostatic scalar potential is a pragmatic
idealization - it is not fundamental and is useful simply because
in many practical circumstances, the time scales of interest are sufficiently
small that charges can be considered stationary in appropriately chosen
inertial frames. Classical electrodynamics subsumes electrostatics
as a special limiting case and so inherits the same pragmatic idealism.
However, as with electrostatics, it is manifestly the case that the
assumption of this idealism does not lead to the classical theory
being in any obvious conflict with our ordinary experience of electrodynamical
phenomenology - but there are no guarantees that this remains the
case in the many extreme electrodynamical regimes probed by modern
science.

The immediate objection to this line of argument is that the canonical
covariant formulation of electromagnetism in terms of the four-vector
potential rests irreducibly upon the assumed existence of the electrostatic
scalar potential - it seems that to deny the existence of this latter
potential is tantamount to denying the existence of a covariant formulation
of electromagnetism. However, in this paper, we show the surprising
result that classical electromagnetism can be reformulated in a relativistically
invariant way without the electrostatic field - that is, purely in
terms of a three-component \emph{relativistic magnetic potential}.
This is extremely interesting of itself but, more significantly, the
process of reformulation reveals a condition of relativistic symmetry
according to which the classical electromagnetic field is one component
of a duality, the second component of which, $G_{ab}$ say, is defined
in terms of a massive vector potential. Further analysis shows that
$G_{ab}$ can be consistently understood in terms of a re-interpretation
of the Lorentz force as a local action contact force. None of this
entails any cost to our ordinary applications of electrodynamics,
but there are clearly fundamental ramifications to our deeper understanding
of the basic phenomenology. In the following, we give a brief overview
of the main body of this paper.

\subsection{The primary question\label{SubSec1.1}}

\emph{Is it possible to have an electrodynamics expressed purely in
terms of some form of (three-component) relativistic magnetic potential
- that is, without} \emph{the electrostatic scalar potential?} 

Once the question has been posed, the key (and apparently trivial)
step is to reformulate the standard definition of the electromagnetic
field tensor, given by \[
F_{ab}=\frac{\partial\phi_{b}}{\partial x^{a}}-\frac{\partial\phi_{a}}{\partial x^{b}},\]
 as\[
F_{ab}=\left\{ \delta_{b}^{r}\frac{\partial}{\partial x^{a}}-\delta_{a}^{r}\frac{\partial}{\partial x^{b}}\right\} \phi_{r}\equiv\mathcal{P}_{ab}^{r}\phi_{r}.\]
 From here, as we show in \S\ref{sec:2}, it is a short step to show
that the primary question, above, receives an unambiguously positive
answer. The result relies upon an analysis of the set of operators
$\left\{ \mathcal{P}_{ab}^{r},\, r=1..4\right\} $ and the realization
that they form a relativistically invariant linear space spanned by
any \emph{three} of $\left\{ \mathcal{P}_{ab}^{r},\, r=1..4\right\} $
- this linear space is defined over the field of differential operations,
$\partial/\partial x^{a},\, a=1..4$, which are assumed to be commutative
on the functions of interest, and with integration as the multiplicative
inverse. Because the invariant linear space contains three independent
elements which are each skew-symmetric with respect to the indices
$\left(a,b\right)$, we denote it as $\mathcal{A}_{sk,3}$ and we
show in \S\ref{sec:2} that its three-dimensionality is the crucial
property from which the idea of the relativistic magnetic potential
flows naturally.

\subsection{The secondary questions\label{sub:1.2}}

The fact that it is possible to reformulate classical electromagnetism
purely in terms of a relativistic magnetic potential in such a simple
way is surprising and suggests, \emph{of itself,} that there may be
further significant consequences attendant upon this changed point
of view.

The key step for progress is the recognition that $\mathcal{A}_{sk,3}$
is actually an invariant subspace of a much bigger sixteen-dimensional
linear space, $\mathcal{S}_{16}$ say, similarly defined over the
field of differential operations (as above). This realization prompted
certain secondary questions:
\begin{quote}
\emph{Given that $\mathcal{S}_{16}$ is spanned by a non-trivial set
of relativistically invariant subspaces, of which one is already known
to be $\mathcal{A}_{sk,3}$, then}  
\begin{enumerate}
\item \emph{what are the remaining invariant spanning subspaces?} 
\item \emph{and what physical significance do they have?}  
\end{enumerate}
\end{quote}
These questions are partly answered by a direct analysis of the structure
of Maxwell's equations themselves, which shows that they rest fundamentally
upon orthogonality properties defined between $\mathcal{A}_{sk,3}$
and two other relativistically invariant subspaces of $\mathcal{S}_{16}$
- a one-dimensional subspace which we label as $\mathcal{B}_{sy,1}$
and a three-dimensional subspace which we label as $\mathcal{C}_{sk,3}$.
The simple fact that the seven-dimensional invariant subspace $\mathcal{A}_{sk,3}\cup\mathcal{B}_{sy,1}\cup\mathcal{C}_{sk,3}$
of $\mathcal{S}_{16}$ is directly implicated in the structure of
electromagnetism then raises automatic questions about the remaining
nine-dimensional subspace of $\mathcal{S}_{16}$ which is so far unaccounted
for.

The complete answer to the first of the secondary questions is found
by representing $\mathcal{S}_{16}$ as a full-rank $16\times16$ relativistically
invariant symmetric matrix, $\mathcal{M}_{16}$ say, defined over
the field of differential operations (as above) and performing an
eigenvalue decomposition of $\mathcal{M}_{16}$. We find that there
are two possibilities for $\mathcal{M}_{16}$, one trivial (diagonal),
and one non-trivial. Since there is only one non-trivial possibility
for $\mathcal{M}_{16}$, then its invariant decomposition provides
a unique invariant decomposition of $\mathcal{S}_{16}$ also. We find
that there are only three distinct eigenvalues of $\mathcal{M}_{16}$,
these being characterized by the parameter values $\lambda=0$ (occurring
nine times), $\lambda=1$ (occurring six times) and $\lambda=2$ (occurring
once only), with associated eigenspaces of dimensions \emph{nine,
six} and \emph{one} respectively, labelled (for convenience here)
as $\mathcal{S}_{9}$, $\mathcal{S}_{6}$ and $\mathcal{S}_{1}$.
Those invariant subspaces already deduced directly from the structure
of classical electromagnetism are now recovered as components in this
decomposition of $\mathcal{M}_{16}$ according to $\mathcal{A}_{sk,3}\in\mathcal{S}_{6}$,
$\mathcal{B}_{sy,1}\equiv\mathcal{S}_{1}$ and $\mathcal{C}_{sk,3}\in\mathcal{S}_{9}$.
The remaining invariant subspaces are recovered similarly according
to $\mathcal{A}_{sy,3}\in\mathcal{S}_{6}$ and $\mathcal{C}_{sy,6}\in\mathcal{S}_{9}$
respectively.

In summary, the relativistically invariant decomposition of $\mathcal{S}_{16}$
is given as $\mathcal{S}_{16}\equiv\mathcal{S}_{1}\cup\mathcal{S}_{6}\cup\mathcal{S}_{9}$
where, in turn:\[
\mathcal{S}_{1}\equiv\mathcal{B}_{sy,1},\,\,\,\,\,\mathcal{S}_{6}\equiv\left\{ \mathcal{A}_{sk,3}\cup\mathcal{A}_{sy,3}\right\} ;\,\,\,\,\,\,\mathcal{S}_{9}\equiv\left\{ \mathcal{C}_{sk,3}\cup\mathcal{C}_{sy,6}\right\} \]
 and where\[
\mathcal{B}_{sy,1}\equiv\left\{ \mathcal{U}_{ab}^{1}\right\} ,\,\,\mathcal{A}_{sk,3}\equiv\left\{ \mathcal{U}_{ab}^{k},\, k=2..4\right\} ,\,\,\mathcal{A}_{sy,3}\equiv\left\{ \mathcal{U}_{ab}^{k},\, k=5..7\right\} ,\]
 \[
\mathcal{C}_{sk,3}\equiv\left\{ \mathcal{U}_{ab}^{k},\, k=8..10\right\} ,\,\,\mathcal{C}_{sy,6}\equiv\left\{ \mathcal{U}_{ab}^{k},\, k=11..16\right\} ,\]
 where each of $\mathcal{U}_{ab}^{k}\in\mathcal{S}_{16},\, k=1..16$,
is a $16\times1$ column, the elements of which are partial differential
operators or zeros.

\subsection{Insight into Maxwell's equations}

Briefly, the invariant decomposition of $\mathcal{S}_{16}$ and a
comparison of this decomposition with the structure of Maxwell's equations,
leads to the ready recognition that the complete \emph{formal} solution
space of these equations consists of second-order tensor objects defined
on the invariant subspaces $\left\{ \mathcal{A}_{sk,3}\cup\mathcal{A}_{sy,3}\right\} \cup\mathcal{C}_{sy,6}$;
that is, general solutions of Maxwell's equations have the form\[
\Delta_{ab}=\left\{ \sum_{k=2}^{4}\mathcal{U}_{ab}^{k}\alpha_{k}(x)+\sum_{k=5}^{7}\mathcal{U}_{ab}^{k}\alpha_{k}(x)\right\} +\sum_{k=11}^{16}\mathcal{U}_{ab}^{k}\alpha_{k}(x),\]
 for potentials, $\alpha_{k}(x)$, and the operators $\mathcal{U}_{ab}^{k}$
being individual elements in the various invariant subspaces. But,
when we consider that the classical electromagnetic field tensor is
defined purely by the three elements of $\mathcal{A}_{sk,3}$ acting
on the components of the relativistic magnetic potential, $\textbf{A}\equiv\left(\alpha_{2},\alpha_{3},\alpha_{4}\right)$
(cf: the considerations of \S\ref{SubSec1.1}), then we see very
clearly that Maxwell's equations are insufficiently constraining for
the purposes of describing purely electromagnetic phenomena. In practice,
of course, as a matter of \emph{diktat} only tensor objects formed
on $\mathcal{A}_{sk,3}$ are ever considered as solutions to Maxwell's
equations - but this does not change the fact that the formal solution
space of these equations is formed on $\left\{ \mathcal{A}_{sk,3}\cup\mathcal{A}_{sy,3}\right\} \cup\mathcal{C}_{sy,6}$.

Given this circumstance, it is illuminating to ask how the solution
space can be further constrained. In considering this question, the
invariant decomposition of $\mathcal{S}_{16}$ makes it clear that,
rather than \emph{selecting} tensor objects that are valid descriptions
of the electromagnetic field, Maxwell's equations, \begin{eqnarray}
\,\,\,\,\,\,\,\,\,\,\,\,\,\,\,\,\,\,\,\,\,\,\,\,\,\,\,\,\,\,\,\,\,\,\frac{\partial F_{\, a}^{i}}{\partial x^{i}} & = & \frac{4\pi}{c}J_{a}\,\label{Intro5}\\
\frac{\partial F_{st}}{\partial x^{r}}+\frac{\partial F_{tr}}{\partial x^{s}}+\frac{\partial F_{rs}}{\partial x^{t}} & = & 0,\label{Intro6}\end{eqnarray}
 actually work by \emph{rejecting} tensor objects that \emph{cannot}
be such valid descriptions. Thus, as we shall show in \S\ref{sub4.3},
condition (\ref{Intro5}) filters out tensor objects formed on the
invariant subspace $\mathcal{B}_{sy,1}$ whilst (\ref{Intro6}) filters
out such objects formed on $\mathcal{C}_{sk,3}$. The remaining admissible
objects are all tensors that can be formed on the invariant subspace
$\left\{ \mathcal{A}_{sk,3}\cup\mathcal{A}_{sy,3}\right\} \cup\mathcal{C}_{sy,6}$.
Given this reality, two obvious points arise:
\begin{enumerate}
\item The invariant decomposition of $\mathcal{S}_{16}$ tells us that the
invariant subspaces $\mathcal{C}_{sk,3}$ and $\mathcal{C}_{sy,6}$
are the dual spanning components of the $\lambda=0$ invariant subspace,
$\mathcal{S}_{9}$, of $\mathcal{S}_{16}$ - they are paired by a
condition of relativistic symmetry. Thus, since (\ref{Intro6}) filters
out all tensor objects formed on $\mathcal{C}_{sk,3}$ as superfluous
for electromagnetism, then relativistic symmetry suggests that, for
reasons of consistency and completeness, Maxwell's equations should
be augmented by that condition which filters out all tensor objects
formed on $\mathcal{C}_{sy,6}$ also. This condition is easily shown
to be\[
\frac{\partial^{2}F_{ss}}{\partial x^{r}\partial x^{r}}-\frac{\partial^{2}F_{rs}}{\partial x^{s}\partial x^{r}}-\frac{\partial^{2}F_{sr}}{\partial x^{r}\partial x^{s}}+\frac{\partial^{2}F_{rr}}{\partial x^{s}\partial x^{s}}=0;\]

\item But, just as $\mathcal{C}_{sk,3}$ and $\mathcal{C}_{sy,6}$ are the
dual spanning components of the $\lambda=0$ subspace, $\mathcal{S}_{9}$,
then so are $\mathcal{A}_{sk,3}$ and $\mathcal{A}_{sy,3}$ the dual
spanning components of the $\lambda=1$ subspace, $\mathcal{S}_{6}$
and must likewise be treated consistently for reasons of relativistic
symmetry. In other words, since $\mathcal{A}_{sk,3}$ is necessary
for the description of electromagnetism then, in some way, so is $\mathcal{A}_{sy,3}$.
The question is \emph{how?}  
\end{enumerate}
Subsequent analysis shows that the potential field upon which $\mathcal{A}_{sy,3}$
must act is a massive vector field which, we are able to show, can
be consistently interpreted as providing a classical \emph{contact
reaction} for the action of the Lorentz force - in other words, the
Lorentz force becomes necessarily re-interpreted as a \emph{contact
force} rather than as a retarded at-a-distance force of the canonical
interpretation. It is this particular consequence which leads necessarily
to a changed view of the nature of the electromagnetic field and to
a potentially deeper understanding of loss mechanisms in electrodynamics.

\subsection{Summary }

By raising a question about the status of the electrostatic field
as a pragmatic idealization, we are able easily to show that classical
electromagnetism can be simply reformulated in terms of a (three-component)
\emph{relativistic} magnetic potential. This reformulation reveals,
via a straightforward analysis, a relativistic symmetry principle
according to which the relativistic magnetic potential is one half
of a duality; the second half, a three-component potential also, has
the property of mass and its presence has the effect of forcing a
natural re-interpretation of the Lorentz force as a local action contact
force. At the level of classical electrodynamics, one unavoidable
consequence of this re-interpretation is a changed understanding of
electrodynamic loss-mechanisms - the processes by which electrodynamic
energy is converted, finally, to thermodynamic energy. Beyond the
concerns of classical electrodynamics, it also becomes clear that
there must be consequences for quantum electrodynamics also, since
this latter theory subsumes the electrostatic field into its basic
structure.

\subsection{Notation notes}
\begin{enumerate}
\item We use the convention that $(x^{1},x^{2},x^{3})$ represent the spatial
axes and $x^{4}\equiv ict$ represents the temporal one with a correspondingly
consistent notation for the four-vector current, $J_{a}$, and the
electromagnetic field tensor, $F_{ab}$; 
\item The metric tensor is now consistently represented by the kronecker-delta,
$\delta_{ab}$; 
\item Although not strictly necessary when using an orthonormal basis, we
make a strict distinction between covariant and contravariant objects
in the conventional way; 
\item We use bold-face to represent an ordinary three-vector; for example,
the position vector $\textbf{x}\equiv(x^{1},x^{2},x^{3})$; 
\item We denote a space-time coordinate as $x\equiv(x^{1},x^{2},x^{3},x^{4})$
or $x\equiv(\textbf{x},ict)$ according to convenience; 
\item For convenience, we use the notation $\partial/\partial x^{a}\equiv X_{a}$,
$\partial/\partial x_{a}\equiv X^{a}$ etc in various places; 
\item By $1/X_{a}$ we mean $\int\cdot dx^{a}$. 
\end{enumerate}

\section{A novel formalism for the electromagnetic field tensor\label{sec:2}}

Static charge does not exist in nature and, in consequence, the assumption
of the existence of the electrostatic field amounts to the assumption
of a certain pragmatic idealism. In the following, we show that this
latter assumption can be removed from the discourse of classical electromagnetism
in an almost trivial way.

We begin by showing how it is possible to restructure the classical
formalism into a new relativistically invariant form which has a \emph{three-component
potential,} $\mathbf{\textbf{A}}\equiv\left(A_{1},A_{2},A_{3}\right)$
say, as its primary object. This potential plays the role of a classical
magnetic vector potential in \emph{any given} frame and is such that
$\textbf{A}\rightarrow\textbf{A}'$ (another three-component potential)
under arbitrary Poincare transformations; for these reasons (and noting
that it is not actually a classical vector) we refer to $\textbf{A}$
as the \emph{relativistic magnetic potential,} or the \emph{RMP.}

\subsection{The details\label{sub2.1}}

In the classical formalism, the electromagnetic field tensor, $F_{ab}$,
is defined in terms of the four-vector potential, \emph{$\phi_{r}$,}
according to\[
F_{ab}=\frac{\partial\phi_{b}}{\partial x^{a}}-\frac{\partial\phi_{a}}{\partial x^{b}}\equiv X_{a}\phi_{b}-X_{b}\phi_{a}\]
 in an obvious notation. This can be reformulated as\begin{equation}
F_{ab}=\mathcal{P}_{ab}^{r}\phi_{r}\equiv\left\{ X_{a}\delta_{b}^{r}-X_{b}\delta_{a}^{r}\right\} \phi_{r}\label{Intro1}\end{equation}
 where the operator $\mathcal{P}_{ab}^{r}$ transforms as a third-order
mixed tensor. Assuming that $X_{a}X_{b}\equiv X_{b}X_{a}$ on the
functions of interest, the significance of this formal change arises
from the fact that \begin{equation}
X_{r}\mathcal{P}_{ab}^{r}\equiv X_{1}\mathcal{P}_{ab}^{1}+X_{2}\mathcal{P}_{ab}^{2}+X_{3}\mathcal{P}_{ab}^{3}+X_{4}\mathcal{P}_{ab}^{4}\equiv0\label{Intro2}\end{equation}
 as is trivially shown. In particular, assuming commutivity etc, this
gives the equivalence\[
\mathcal{P}_{ab}^{4}\equiv-\left\{ \mathcal{P}_{ab}^{1}\frac{X_{1}}{X_{4}}+\mathcal{P}_{ab}^{2}\frac{X_{2}}{X_{4}}+\mathcal{P}_{ab}^{3}\frac{X_{3}}{X_{4}}\right\} \]
 so that $F_{ab}$ at (\ref{Intro1}) can be equivalently expressed
as\begin{equation}
F_{ab}=\sum_{r=1}^{3}\mathcal{P}_{ab}^{r}\left(\phi_{r}-\frac{X_{r}}{X_{4}}\phi_{4}\right)\equiv\sum_{r=1}^{3}\mathcal{P}_{ab}^{r}A_{r}\label{Intro3}\end{equation}
 in an obvious notation.

\subsection{Relativistic invariance of (\ref{Intro3})\label{sub:2.2}}

The structure of (\ref{Intro3}) is relativistically invariant: to
see this, we expand (\ref{Intro3}) explicitly to get\begin{equation}
\left\{ F_{ab}\right\} =\left(\begin{array}{cccc}
0 & X_{1}A_{2}-X_{2}A_{1} & X_{1}A_{3}-X_{3}A_{1} & -X_{4}A_{1}\\
 & 0 & X_{2}A_{3}-X_{3}A_{2} & -X_{4}A_{2}\\
 &  & 0 & -X_{4}A_{3}\\
 &  &  & 0\end{array}\right)\label{Intro4}\end{equation}
 so that, in any given frame, $\textbf{A}\equiv\left(A_{1},A_{2},A_{3}\right)$
plays the part of the classical magnetic vector potential in a frame
for which the electrostatic scalar potential is explicitly absent.
Consequently, in this frame the four-vector potential of canonical
theory can be written as \[
\Phi\equiv\left(\Phi_{1},\Phi_{2},\Phi_{3},\Phi_{4}\right)=\left(A_{1},A_{2},A_{3},0\right)\]
 so that, by (\ref{Intro1}) and (\ref{Intro3}), we have\[
F_{ab}=\mathcal{P}_{ab}^{r}\Phi_{r}=\sum_{r=1}^{3}\mathcal{P}_{ab}^{r}A_{r}.\]
 Consequently, under the orthogonal transformation $x\rightarrow\hat{x}=T\left(x-x_{0}\right)$,
we have\begin{eqnarray*}
\hat{F}_{ab} & = & T_{a}^{i}T_{b}^{j}F_{ij}=T_{a}^{i}T_{b}^{j}\mathcal{P}_{ij}^{r}\Phi_{r}=T_{a}^{i}T_{b}^{j}\left(X_{i}\delta_{j}^{r}-X_{j}\delta_{i}^{r}\right)\Phi_{r}\\
 & = & \left(\hat{X}_{a}T_{b}^{r}-\hat{X}_{b}T_{a}^{r}\right)\Phi_{r}=\hat{X}_{a}\hat{\Phi}_{b}-\hat{X}_{b}\hat{\Phi}_{a}\\
 & = & \hat{\mathcal{P}}_{ab}^{r}\hat{\Phi}_{r}=\sum_{r=1}^{3}\hat{\mathcal{P}}_{ab}^{r}\left(\hat{\Phi}_{r}-\frac{\hat{X}_{r}}{\hat{X}_{4}}\hat{\Phi}_{4}\right)\equiv\sum_{r=1}^{3}\hat{\mathcal{P}}_{ab}^{r}\hat{A}_{r}\,.\end{eqnarray*}
 Thus, $F_{ab}\rightarrow\hat{F}_{ab}$ under $x\rightarrow\hat{x}=T(x-x_{0})$
also implies\[
\sum_{r=1}^{3}\mathcal{P}_{ab}^{r}A_{r}\rightarrow\sum_{r=1}^{3}\hat{\mathcal{P}}_{ab}^{r}\hat{A}_{r}\]
 so that the three-component representation of electromagnetism, given
at (\ref{Intro3}), is relativistically invariant as stated. In other
words, the electrostatic scalar potential is eliminated from the discourse
of electromagnetism. For this reason, we shall refer to $\textbf{A}$
as the \emph{relativistic magnetic potential} or RMP.

\subsection{Explicit transformation law for the relativistic magnetic potential\label{sub2.3}}

The transformation law for the RMP is implicit from the considerations
of \S\ref{sub:2.2} above. Specifically, the process has been $\textbf{A}\rightarrow\Phi\rightarrow\hat{\Phi}\rightarrow\hat{\textbf{A}}$
so that, under the coordinate transformation $x\rightarrow\hat{x}=T\left(x-x_{0}\right)$,
the RMP transforms according to\begin{equation}
A_{r}\rightarrow\hat{A}_{r}\equiv\left(\hat{\Phi}_{r}-\frac{\hat{X}_{r}}{\hat{X}_{4}}\hat{\Phi}_{4}\right),\,\, r=1,2,3\label{eqn28f}\end{equation}
 where each of $\hat{\Phi}_{k},\, k=1..4$ is just a scrambling of
the three components of $\textbf{A}$.

\subsection{Gauge properties of the RMP formalism\label{sub2.4}}

Canonical electromagnetism, expressed in terms of the four-vector
potential, $\phi_{r}$, is invariant under the gauge transformation
$\phi_{r}'\rightarrow\phi_{r}+X_{r}\psi$ where $\psi$ is an arbitrary
differentiable function of $x\equiv\left(\textbf{x},ct\right)$. It
is interesting to note that, \emph{in any given frame}, the relativistically
invariant RMP formalism is recovered exactly from the canonical formalism
by applying the temporal gauge, \[
X_{4}\psi=-\phi_{4}\]
since this implies that\[
\phi_{r}'\rightarrow\left(\phi_{r}-\frac{X_{r}}{X_{4}}\phi_{4}\right),\,\, r=1,2,3\]
together with $\phi'_{4}=0$, which is to be compared with (\ref{eqn28f}).
For this reason, the temporal gauge can be considered as the natural
\emph{relativistic} gauge. \\
\\
However, even though the RMP formalism can be recovered from the
canonical four-vector formalism by the application of the temporal
gauge, it is still the case, as (\ref{Intro4}) makes plain, that
the RMP formalism is invariant under the additional gauge transformation
$\textbf{A}'\rightarrow\textbf{A}+\nabla\psi$ where $\psi$ is an
arbitrary differentiable function of $\textbf{x}$ \emph{only.} Thus,
the RMP formalism is only defined to within $\nabla\psi$, and needs
an additional condition on $\textbf{A}$ to uniquely specify it. \\
\\
The fact that the RMP formalism is equivalent to the canonical
four vector formalism in any given frame under the application of
a specific gauge indicates that we should expect to recover all the
important predictions of classical QED from a quantized field theory
based directly upon the RMP formalism - but we might expect changes
of interpretation. We shall discuss the interpretation issues in $\S$\ref{sec:6}.

\subsection{Immediate implications\label{sub:2.5}}

The operators $\mathcal{P}_{ab}^{r}\equiv\left\{ X_{a}\delta_{b}^{r}-X_{b}\delta_{a}^{r}\right\} ,\, r=1..4$
represent {}``hidden structure'' in electromagnetism and in the
following we show that they lie in an invariant subspace of a much
bigger linear space. This fact suggest the existence of further physically
significant hidden structure within electromagnetism.

$\mathcal{P}_{ab}^{r}$ has the following properties:
\begin{itemize}
\item it transforms as a mixed third-order tensor under the Poincare group; 
\item it can be interpreted as a set of four $(r=1..4)$ columns defined
in a sixteen dimensional column space, $\left(a,b\right)=(1,1),(1,2)...(4,4)$; 
\item assuming that $X_{a}X_{b}\equiv X_{b}X_{a}$ over the functions of
interest, then $X_{k}\mathcal{P}_{ab}^{k}=0$. Thus,  each of the
four $16\times1$ columns, $\left\{ \mathcal{P}_{ab}^{r},\, r=1..4\right\} $,
can be considered as a vector in a relativistically invariant linear
space which is spanned by any three of $\left\{ \mathcal{P}_{ab}^{r},\, r=1..4\right\} $
- this linear space is defined over the field of differential operations,
assumed commutative, which has integration as the multiplicative inverse. 
\end{itemize}
For convenience, therefore, we:
\begin{itemize}
\item denote the sixteen-dimensional column space as $\mathcal{S}_{16}$; 
\item denote the invariant subspace spanned by any three of $\left\{ \mathcal{P}_{ab}^{r},\, r=1..4\right\} $
as $\mathcal{A}_{sk,3}\in\mathcal{S}_{16}$. The subscript $sk$ denotes
the skew-symmetry of the $\mathcal{P}_{ab}^{r}$ with respect to interchange
of $a$ and $b$. 
\end{itemize}
Since $\mathcal{A}_{sk,3}$ has fundamental significance to electromagnetism,
the obvious questions are:
\begin{enumerate}
\item \emph{Given that $\mathcal{S}_{16}$ is spanned by a set of non-trivially
defined relativistically invariant subspaces, of which one is already
known to be $\mathcal{A}_{sk,3}$, what are the remaining relativistically
invariant subspaces spanning $\mathcal{S}_{16}$?} 
\item \emph{What do these invariant subspaces represent?} 
\end{enumerate}

\section{Hidden structure in the classical equations\label{sec:Hidden}}

An initial insight into the invariant decomposition of $\mathcal{S}_{16}$
can be obtained by considering Maxwell's equations explicitly: when
expressed in terms of the field tensor, the microscopic form in the
presence of conserved current $J$ is conventionally written \begin{equation}
\frac{\partial F_{a}^{i}}{\partial x^{i}}=\frac{4\pi}{c}J_{a}\,\longrightarrow\,\frac{\partial^{2}F_{ij}}{\partial x_{i}\partial x_{j}}=0\label{eqn.Intro1}\end{equation}
 together with \begin{equation}
\frac{\partial F_{st}}{\partial x^{r}}+\frac{\partial F_{tr}}{\partial x^{s}}+\frac{\partial F_{rs}}{\partial x^{t}}=0.\label{eqn.Intro2}\end{equation}
 To within the numerical factor, we can consider the second of (\ref{eqn.Intro1})
together with (\ref{eqn.Intro2}) to represent Maxwell's equations
which, in terms of $F_{ab}\equiv\mathcal{P}_{ab}^{k}\phi_{k}$, can
be formally expressed as\begin{eqnarray}
\,\,\,\,\,\,\,\,\,\,\,\,\,\,\,\,\,\,\,\,\,\,\,\,\,\,\,\,\,\,\,\,\,\,\frac{\partial^{2}F_{ij}}{\partial x_{i}\partial x_{j}} & \equiv & \mathcal{N}^{ij}\mathcal{P}_{ij}^{k}\phi_{k}=0\nonumber \\
\frac{\partial F_{st}}{\partial x^{r}}+\frac{\partial F_{tr}}{\partial x^{s}}+\frac{\partial F_{rs}}{\partial x^{t}} & \equiv & \mathcal{R}_{rst}^{ij}\mathcal{P}_{ij}^{k}\phi_{k}=0,\label{eqn.Intro6}\end{eqnarray}
 respectively, for linear differential operators $\mathcal{N}^{ab}$
and $\mathcal{R}_{rst}^{ab}$. Since an entirely \textit{\emph{arbitrary}}
definition of the potentials $(\phi_{1},\phi_{2},\phi_{3},\phi_{4})$
satisfies (\ref{eqn.Intro6}), it follows that $\mathcal{N}^{ij}\mathcal{P}_{ij}^{k}\equiv0$
and $\mathcal{R}_{rst}^{ij}\mathcal{P}_{ij}^{k}\equiv0$; that is,
the canonical covariant form of Maxwell's equations can be considered
based on algebraic orthogonality properties between $\mathcal{P}_{ab}^{r}$,
$\mathcal{R}_{rst}^{ab}$ and $\mathcal{N}^{ab}$.

\subsection{$\mathcal{N}_{ab}$ and $\mathcal{R}_{ab}^{rst}$ as elements in
invariant subspaces of $\mathcal{S}_{16}$\label{sub3.1}}

From the first of (\ref{eqn.Intro6}), we have $\mathcal{N}_{ab}\equiv X_{a}X_{b}$
which transforms as a second-order covariant tensor and which, considered
as a $16\times1$ column, is a relativistically invariant one-dimensional
subspace of $\mathcal{S}_{16}$. Since $\mathcal{N}_{ab}$ is symmetric
in $a$ and $b$ then we denote the invariant subspace represented
by $\mathcal{N}_{ab}$ as $\mathcal{B}_{sy,1}$.

From the second of (\ref{eqn.Intro6}), it is clear that $\mathcal{R}_{ab}^{rst}$
must transform as a mixed fifth-order tensor and its explicit form
can be deduced to be given by \[
\mathcal{R}_{ab}^{rst}\equiv\frac{X^{r}X^{s}X^{t}}{2\, X^{a}X^{b}}\left(\left(\delta_{a}^{r}-\delta_{a}^{s}\right)\left(\delta_{b}^{s}-\delta_{b}^{t}\right)-\left(\delta_{b}^{r}-\delta_{b}^{s}\right)\left(\delta_{a}^{s}-\delta_{a}^{t}\right)\right).\]
 Furthermore, since it is easily shown that\[
X^{1}\mathcal{R}_{ab}^{234}-X^{2}\mathcal{R}_{ab}^{341}+X^{3}\mathcal{R}_{ab}^{412}-X^{4}\mathcal{R}_{ab}^{123}\equiv0\]
 then $\mathcal{R}_{ab}^{rst}$ gives rise to only three independent
$16\times1$ columns each of which is also skew-symmetric with respect
to $a$ and $b$. For this reason, we refer to the invariant subspace
spanned by these columns as $\mathcal{C}_{sk,3}$.

\selectlanguage{english}%

\subsection{Wave types supported by the Relativistic Magnetic Potential}

\label{subsec.Waves} It is shown that, according to RMP electrodynamics,
wavy solutions for the RMP are composed of two distinct kinds of wave:
the first kind is a propagating transverse wave, whilst the second
kind, which is novel, is a stationary longitudinal wave. It is shown
that the propagating transverse component corresponds identically
to those solutions which arise in the conventional formalism when
the temporal gauge is used. The stationary longitudinal component
has no counterpart in the conventional formalism.

Using the notation $\textbf{A}\equiv(A_{1},A_{2},A_{3})$ in (\ref{Intro3}),
then (\ref{eqn.Intro1}) can be written: \[
\frac{\partial F_{a}^{i}}{\partial x^{i}}\equiv\sum_{r=1}^{3}X_{i}\left(X_{a}\delta_{r}^{i}-X^{i}\delta_{ra}\right)A_{r}=\frac{4\pi}{c}J_{a}\]
 which - upon remembering $X_{a}\equiv\partial/\partial x^{a}$ -
can be written as the two equations \begin{eqnarray}
\Box\textbf{A}-\nabla\left(\nabla\cdot\textbf{A}\right)=-\frac{4\pi}{c}\textbf{J}\label{eqn.Matvac1}\\
\frac{\partial}{\partial x^{4}}\left(\nabla\cdot\textbf{A}\right)=\frac{4\pi}{c}J_{4}.\nonumber \end{eqnarray}
 Given ${\textbf{J}}$, and hence ${\textbf{A}}$ via (\ref{eqn.Matvac1}),
the second of these equations provides a \textit{definition} of $J^{4}$.
Consider now, a wave given by \[
\textbf{A}_{wave}=\textbf{A}_{0}\exp(i\textbf{n}\cdot\textbf{x}),\]
 where $\textbf{n}\equiv(n_{1},n_{2},n_{3},n_{4})$, $\textbf{x}\equiv(x_{1},x_{2},x_{3},x_{4})$
and $\textbf{A}_{0}$ is a constant three-vector. The requirement
that $\textbf{A}_{wave}$ satisfies (\ref{eqn.Matvac1}) with $\textbf{J}=0$
leads to the system of equations \begin{equation}
(\mathbf{n\cdot n})\mathbf{A}_{0}=(\mathbf{\hat{n}}\mathbf{\cdot A}_{0})\mathbf{\hat{n}}\label{eqn.Waves1}\end{equation}
 where $\mathbf{\hat{n}}\equiv(n_{1},n_{2},n_{3})$ and, from this,
we can form the scalar equation \begin{equation}
(\mathbf{n\cdot n})(\mathbf{\hat{n}}\mathbf{\cdot A}_{0})=(\mathbf{\hat{n}}\mathbf{\cdot A}_{0})(\mathbf{\hat{n}\cdot\hat{n}}).\label{eqn.Waves2}\end{equation}
 This latter equation has two possible solutions which, together,
form a basis for the general solution of (\ref{eqn.Matvac1}):

\subsubsection*{Case 1: The Transverse Wave: $\mathbf{\hat{n}\cdot A}_{0}=0$}

In this case, (\ref{eqn.Waves1}) only has a non-trivial solution
if $\mathbf{n\cdot n}=0$. Consequently, this solution is given by
\begin{equation}
\mathbf{A}_{T}=\mathbf{A}_{0}\exp(i\mathbf{n\cdot x}),\,\,\,\mathbf{n\cdot n}=0,\,\,\,\mathbf{\hat{n}\cdot A}_{0}=0,\label{eqn.Waves2a}\end{equation}
 which corresponds to a \textit{transverse} wave propagating with
speed $c$.

\subsubsection*{Case 2: The Longitudinal Wave: $\mathbf{\hat{n}\cdot A}_{0}\neq0$}

In this case, (\ref{eqn.Waves2}) gives $\mathbf{n\cdot n}=\mathbf{\hat{n}\cdot\hat{n}}$,
and this can only be true if $n_{4}=0$. From (\ref{eqn.Waves1})
we now get the equation \[
(\mathbf{\hat{n}\cdot\hat{n}})\mathbf{A}_{0}=(\mathbf{\hat{n}\cdot A}_{0})\mathbf{\hat{n}}\]
 which is easily seen to have the solution $\mathbf{A}_{0}=\alpha\mathbf{\hat{n}}$
for arbitrary $\alpha$. To summarize, this solution is given by \begin{equation}
\mathbf{A}_{L}=\alpha\mathbf{\hat{n}}\exp(i\mathbf{\hat{n}\cdot\hat{x}})\label{eqn.Waves3}\end{equation}
 where $\mathbf{\hat{x}}=(x_{1},x_{2},x_{3})$, and this corresponds
to a \textit{longitudinal} stationary wave. This wave is easily shown
to give $\textbf{E}=\textbf{B}=0$, so that a non-trivial RMP can
have a zero electromagnetic effect.

To summarize, we arrive at the conclusion that the RMP supports two
kinds of waves in free space: a propagating transverse wave and a
stationary longitudinal wave (which has no counterpart in the conventional
formalism) so that the general wavy solution to the homogeneous form
of (\ref{eqn.Matvac1}) is given by \[
\textbf{A}_{wave}=\textbf{A}_{T}(\textbf{x},ct)+\textbf{A}_{L}(\textbf{x}),\]
 where ${\textbf{A}}_{T}$ is the transverse wave propagating with
speed $c$, and $\textbf{A}_{L}$ is the stationary longitudinal wave.
The component $\textbf{A}_{T}$ gives rise to propagating transverse
electromagnetic fields, and the general phenomenology that, when such
a field is created, any charged particle anywhere will eventually
feel its effect. The component $\textbf{A}_{L}$ gives rise to a \textit{zero}
electromagnetic field ($\textbf{E}=\textbf{B}=0$) and so no electromagnetic
effect at all is propagated; however, the possibility exists that
$\mathbf{A}_{L}$ is implicated as the causative agent at the classical
level of the Aharanov-Bohm effect. 

\selectlanguage{british}%

\section{Representation of $\mathcal{S}_{16}$ as a $16\times16$ invariant
matrix}

An analysis of classical electromagnetism has shown that its properties
are encapsulated in that relativistically invariant subspace of $\mathcal{S}_{16}$
which we have labelled as $\mathcal{A}_{sk,3}$, whilst Maxwell's
equations themselves are statements of the orthogonality properties
$\mathcal{B}_{sy,1}\perp\mathcal{A}_{sk,3}$ and $\mathcal{C}_{sk,3}\perp\mathcal{A}_{sk,3}$
respectively, where $\mathcal{B}_{sy,1}$ and $\mathcal{C}_{sk,3}$
are two other relativistically invariant subspaces of $\mathcal{S}_{16}$.
Thus, since certain invariant subspaces of $\mathcal{S}_{16}$ are
implicated in classical electromagnetism, it becomes natural to query
the structure of the remaining invariant subspaces of $\mathcal{S}_{16}$
and to investigate their potential physical significance. To pursue
these questions, we need a complete basis for $\mathcal{S}_{16}$
in terms of its relativistically invariant subspaces which we propose
to recover in the following way:
\begin{enumerate}
\item represent $\mathcal{S}_{16}$ as a relativistically invariant full-rank
$16\times16$ symmetric matrix, $\mathcal{M}_{16}$; 
\item perform an eigenvalue decomposition of $\mathcal{M}_{16}$. Since
$\mathcal{M}_{16}$ is relativistically invariant by construction,
its eigenspace will necessarily consist of a set of relativistically
invariant subspaces, one for each distinct eigenvalue. This set of
invariant subspaces will provide the required basis for $\mathcal{S}_{16}$. 
\end{enumerate}
Since $\mathcal{M}_{16}$ is an invariant $16\times16$ matrix, we
can deduce that rows $1..16$ can be represented as the covariant
indices $(m,n)=(1,1)..(4,4)$ with the columns $1..16$ then being
represented as the contravariant indices $(r,s)=(1,1)..(4,4)$ ordered
similarly so that $\mathcal{M}_{16}$ can be expressed as the mixed\emph{
fourth-order} tensor $\mathcal{M}_{mn}^{rs}$. We now note that, since
the building blocks of $\mathcal{M}_{mn}^{rs}$ are the second-order
tensor $\delta_{ab}$ and the first-order tensor $X_{a}$, it follows
that the first-order tensor object must necessarily occur in pairs,
similar to $X_{a}X_{b}$. With this understanding, we find only two
possibilities for $\mathcal{M}_{16}\equiv\mathcal{M}_{mn}^{rs}$:\begin{eqnarray}
\mathcal{M}_{mn}^{rs} & \equiv & \sigma_{mn}^{ij:rs}X_{i}X_{j}\equiv\left\{ \delta_{m}^{i}\delta^{jr}\delta_{n}^{s}+\delta_{n}^{i}\delta^{js}\delta_{m}^{r}\right\} X_{i}X_{j}\,,\label{eqn4.1}\\
\nonumber \\\mathcal{M}_{mn}^{rs} & \equiv & \sigma_{mn}^{ij:rs}X_{i}X_{j}\equiv\left\{ \delta_{m}^{r}\delta^{ij}\delta_{n}^{s}+\delta_{n}^{r}\delta^{ij}\delta_{m}^{s}\right\} X_{i}X_{j}\equiv-\left\{ \delta_{m}^{r}\delta_{n}^{s}+\delta_{n}^{r}\delta_{m}^{s}\right\} \Box\nonumber \end{eqnarray}
Of these alternatives, the second is purely diagonal with the d'Alembertian
as the diagonal element and is therefore the trivial case. There is
thus a unique non-trivial alternative, defined at (\ref{eqn4.1}),
so that there will be a uniquely defined set of relativistically invariant
subspaces which span $\mathcal{S}_{16}$.

\subsection{The eigenspace of $\mathcal{M}_{16}$\label{subsec:4.1}}

Since $\mathcal{M}_{16}$ is defined over the field of differential
operations, its eigenvalues are not ordinary numbers but must be multiples
of a relativistically invariant scalar quantity defined on this field
- that is, of $\Box\equiv-X_{i}X^{i}$, the d'Alembertian. Therefore,
the eigenspace of $\mathcal{M}_{16}$, defined as the set of $16\times1$
vectors, $\mathcal{U}^{k},\, k=1..16$ say - where $\mathcal{U}^{1}\equiv\left(\mathcal{U}_{11}^{1},\mathcal{U}_{12}^{1},\mathcal{U}_{13}^{1},...,\mathcal{U}_{44}^{1}\right)^{T}$
etc - can be found by solving the eigenvalue problem $\mathcal{M}_{16}\mathcal{U}^{k}=-\lambda\Box\mathcal{U}^{k}$
which, using (\ref{eqn4.1}), can be written explicitly as \begin{eqnarray}
X_{a}X^{i}\mathcal{U}_{ib}^{k}+X_{b}X^{i}\mathcal{U}_{ai}^{k} & = & -\lambda\Box\mathcal{U}_{ab}^{k},\,\, k=1..16.\label{eqn4a}\end{eqnarray}
 We find only three distinct eigenvalues, $\lambda=0$ (occurring
nine times), $\lambda=1$ (occurring six times) and $\lambda=2$ (occurring
once only), where the corresponding eigenspaces have dimensions nine,
six and one respectively. Thus, we obtain the orthogonal decomposition
$\mathcal{S}_{16}\equiv\mathcal{S}_{1}\cup\mathcal{S}_{6}\cup\mathcal{S}_{9}$
where
\begin{itemize}
\item ${\cal S}_{1}\equiv\left\{ \mathcal{U}^{1}\right\} \equiv\mathcal{B}_{sy,1}$
which is a one-dimensional space corresponding to the eigenvalue $\lambda=2$
and is the invariant subspace already deduced directly from Maxwell's
equations in \S\ref{sub3.1}; 
\item ${\cal S}_{6}\equiv\left\{ \mathcal{U}^{2}..\,\mathcal{U}^{7}\right\} \equiv\mathcal{A}_{sk,3}\cup\mathcal{A}_{sy,3}$
which is a six-dimensional space corresponding to the six-times-occurring
eigenvalue $\lambda=1$.

\begin{itemize}
\item The skew-symmetric component, $\left\{ \mathcal{U}^{2}..\,\mathcal{U}^{4}\right\} \equiv\mathcal{A}_{sk,3}$,
is that already deduced in \S\ref{sub:2.5} to be the basis of the
relativistic magnetic potential; 
\item The associated symmetric component, $\left\{ \mathcal{U}^{5}..\,\mathcal{U}^{7}\right\} \equiv\mathcal{A}_{sy,3}$,
is new and it will be shown that it corresponds to a massive vector
field which can be consistently interpreted as providing a \emph{contact
reaction} for the action of the Lorentz force; 
\end{itemize}
\item ${\cal S}_{9}\equiv\left\{ \mathcal{U}^{8}..\,\mathcal{U}^{16}\right\} \equiv\mathcal{C}_{sk,3}\cup\mathcal{C}_{sy,6}$
which is a nine-dimensional space corresponding to the nine-times-repeated
eigenvalue $\lambda=0$.

\begin{itemize}
\item The skew-symmetric component, $\left\{ \mathcal{U}^{8}..\,\mathcal{U}^{10}\right\} \equiv\mathcal{C}_{sk,3}$,
is that already deduced directly from Maxwell's equations in \S\ref{sub3.1}; 
\item The associated symmetric component, $\left\{ \mathcal{U}^{11}..\,\mathcal{U}^{16}\right\} \equiv\mathcal{C}_{sy,6}$,
is the only one of the invariant subspaces not to have an obvious
role within electrodynamics. 
\end{itemize}
\end{itemize}
The precise algebraic structure of the eigenvectors is given in Appendix
\ref{Appendix:B}.

\subsection{Orthogonality Properties\label{subsec:4.2}}

Since the matrix $\mathcal{M}_{16}$ is symmetric then, as a standard
result, eigenvectors drawn from the distinct eigenspaces of this matrix
are mutually orthogonal in the sense that if vectors $\mathbf{U}\equiv\left(U_{ab}\right)$,
$\mathbf{V}\equiv\left(V_{ab}\right)$ and $\mathbf{W}\equiv\left(W_{ab}\right)$
are such that $\mathbf{U}\in\mathcal{B}_{sy,1}$, $\mathbf{V}\in\left\{ \mathcal{A}_{sk,3}\cup\mathcal{A}_{sy,3}\right\} $
and $\mathbf{W}\in\left\{ \mathcal{C}_{sk,3}\cup\mathcal{C}_{sy,6}\right\} $,
then\[
\mathbf{U}^{T}\mathbf{V}\equiv U^{ij}V_{ij}\equiv0,\,\,\,\,\mathbf{U}^{T}\mathbf{W}\equiv U^{ij}W_{ij}\equiv0,\,\,\,\,\mathbf{V}^{T}\mathbf{W}\equiv V^{ij}W_{ij}\equiv0\]
and is easily confirmed explicitly by reference to Appendix \ref{Appendix:B}.

These orthogonality properties are fundamental to much of what follows.

\subsection{Insight into Maxwell's equations\label{sub4.3}}

Suppose $H_{ab}$ is a tensor object defined on\[
\mathcal{S}_{16}\equiv\mathcal{B}_{sy,1}\cup\left\{ \mathcal{A}_{sk,3}\cup\mathcal{A}_{sy,3}\right\} \cup\left\{ \mathcal{C}_{sk,3}\cup\mathcal{C}_{sy,6}\right\} \]
 and suppose further that $H_{ab}$ must satisfy Maxwell's equations
which, from (\ref{eqn.Intro6}), can be written \begin{eqnarray*}
\mathcal{N}^{ij}H_{ij} & = & 0,\,\,\,\,\mathcal{N}^{ab}\in\mathcal{B}_{sy,1\,,}\\
\mathcal{R}_{rst}^{ij}H_{ij} & = & 0,\,\,\,\,\mathcal{R}_{rst}^{ab}\in\mathcal{C}_{sk,3}\,.\end{eqnarray*}
 Since the invariant subspaces are mutually orthogonal (cf \S\ref{subsec:4.2}),
then this form makes it clear that the first Maxwell equation, above,
is the condition that $H_{ab}$ \emph{cannot} be defined on $\mathcal{B}_{sy,1}$
whilst the second Maxwell equation, above, is the condition that $H_{ab}$
\emph{cannot} be defined on $\mathcal{C}_{sk,3}$. In other words,
Maxwell's equations act to (partially) define the electromagnetic
field tensor by \emph{excluding} certain possibilities. They leave
open the possibility that $H_{ab}$ can be any tensor object defined
on the invariant subspace $\left\{ \mathcal{A}_{sk,3}\cup\mathcal{A}_{sy,3}\right\} \cup\mathcal{C}_{sy,6}$.

At this point, we note that $\mathcal{C}_{sk,3}$ and $\mathcal{C}_{sy,6}$
are the dual spanning components of the $\lambda=0$ subspace, $\mathcal{S}_{9}$,
of $\mathcal{S}_{16}$ - they are paired by relativistic symmetry.
It follows that since the second of the Maxwell equations, above,
filters out solutions defined on $\mathcal{C}_{sk,3}$, then there
should be an augmenting equation designed to filter out solutions
defined on $\mathcal{C}_{sy,6}$ also. The fully augmented system
has the structure \begin{eqnarray}
\mathcal{N}^{ij}H_{ij} & = & 0,\,\,\,\,\mathcal{N}^{ab}\,\in\mathcal{B}_{sy,1}\,,\nonumber \\
\mathcal{R}_{rst}^{ij}H_{ij} & = & 0,\,\,\,\,\mathcal{R}_{rst}^{ab}\in\mathcal{C}_{sk,3}\,,\label{eqn4d}\\
\mathcal{Q}_{rs}^{ij}H_{ij} & = & 0,\,\,\,\,\mathcal{Q}_{rs}^{ab}\,\,\in\mathcal{C}_{sy,6}\,,\nonumber \end{eqnarray}
 where the precise structure of $\mathcal{Q}_{rs}^{ab}$ is given
in appendix \S\ref{sub:B5}. The general solutions of this augmented
system consist of all tensor objects defined on $\left\{ \mathcal{A}_{sk,3}\cup\mathcal{A}_{sy,3}\right\} $
which, from the $\lambda=1$ case of (\ref{eqn4a}), is also the complete
solution space of\begin{equation}
X_{a}X^{i}H_{ib}+X_{b}X^{i}H_{ai}=-\Box H_{ab}\label{eqn:4f}\end{equation}
 so that (\ref{eqn4d}) and (\ref{eqn:4f}) are \emph{exactly equivalent.}
For either system, complete solutions have the general structure\[
H_{ab}\equiv F_{ab}+G_{ab}\equiv\sum_{k=2}^{4}\mathcal{U}_{ab}^{k}\alpha_{k}(x)+\sum_{k=5}^{7}\mathcal{U}_{ab}^{k}\alpha_{k}(x),\,\,\,\,\mathcal{U}_{ab}^{k}\in\left\{ \mathcal{A}_{sk,3}\cup\mathcal{A}_{sy,3}\right\} .\]
 For the strict purpose of defining the electromagnetic field, $F_{ab}$,
in isolation, a solution space consisting of all tensor objects defined
on $\mathcal{A}_{sk,3}$ is sufficient. But relativistic symmetry
tells us that $\mathcal{A}_{sk,3}$ and $\mathcal{A}_{sy,3}$ are
the dual components of the $\lambda=1$ invariant subspace, $\mathcal{S}_{6}$.
In other words, relativistic symmetry requires that, whenever the
electromagnetic field, $F_{ab}$, is present, a second field, $G_{ab}$
defined on $\mathcal{A}_{sy,3}$, must also be present. Thus, the
question becomes: \emph{what does $G_{ab}$ represent?}

\section{A massive vector field from $\mathcal{A}_{sy,3}$\label{sec.vecboson}}

Relativistic symmetry implies that the electromagnetic field, $F_{ab}$,
and the field $G_{ab}$ defined as\[
G_{ab}=\sum_{k=5}^{7}\mathcal{U}_{ab}^{k}\alpha_{k}(x),\,\,\,\,\mathcal{U}_{ab}^{k}\in\mathcal{A}_{sy,3}\]
 for potentials $\textbf{C}\equiv\left(\alpha_{5},\alpha_{6},\alpha_{7}\right)$,
are the components of an irreducible duality. In this section, we
shall show:
\begin{itemize}
\item that $\textbf{C}$ represents a massive vector field; 
\item how $\textbf{C}$ transforms under coordinate transformations $x\rightarrow\hat{x}=T\left(x-x_{0}\right)$;
\item how the individual components of $\textbf{C}$ satisfy the Klein-Gordon
equation.
\end{itemize}
It is argued, \S\ref{sec:6}, that the massive vector field generated
by $\mathcal{A}_{sy,3}$ provides a natural mechanism for electrodynamical
reaction forces, thereby allowing the Lorentz force to be reinterpreted
as a direct local action contact force.

\subsection{Preliminaries\label{sub:5.1}}

\label{subsec.vecboson1} From Appendix \ref{Appendix:B}, a basis
for the invariant subspace $\mathcal{A}_{sy,3}$ is given by\[
\mathcal{U}_{ab}^{k}\equiv X_{a}\left(X^{r}\delta_{b}^{s}-X^{s}\delta_{b}^{r}\right)+X_{b}\left(X^{r}\delta_{a}^{s}-X^{s}\delta_{a}^{r}\right),~~~k=5,6,7\]
 where for $k=(5,6,7)$, then $(r,s)$ is three distinct pairs chosen
from $(1,2,3,4)$. The basis is most conveniently chosen by picking
any one of the four digits and pairing it with the remaining three.
We shall use $(k,r,s)=(5,1,4),(6,2,4),(7,3,4)$ for definiteness.

The most general field which can be formed from the operators lying
in this subspace is given by\begin{eqnarray*}
G_{ab} & = & \sum_{k=5}^{7}\mathcal{U}_{ab}^{k}\alpha_{k}(x)\equiv\sum_{k=5}^{7}\left\{ X_{a}\left(X^{r}\delta_{b}^{s}-X^{s}\delta_{b}^{r}\right)+X_{b}\left(X^{r}\delta_{a}^{s}-X^{s}\delta_{a}^{r}\right)\right\} \alpha_{k}(x),\\
 &  & (k,r,s)=(5,1,4),\,(6,2,4),\,(7,3,4),\end{eqnarray*}
 which, noting that $X^{r}\delta_{t}^{s}-X^{s}\delta_{t}^{r}\equiv\mathcal{P}_{t}^{rs}\in\mathcal{A}_{sk,3}$,
can be written as\begin{eqnarray}
G_{ab} & = & \sum_{k=5}^{7}\mathcal{U}_{ab}^{k}\alpha_{k}(x)\equiv\sum_{k=5}^{7}\left\{ X_{a}\mathcal{P}_{b}^{rs}+X_{b}\mathcal{P}_{a}^{rs}\right\} \alpha_{k}(x).\nonumber \\
 &  & (k,r,s)=(5,1,4),\,(6,2,4),\,(7,3,4).\label{eqn.22a}\end{eqnarray}
 Finally, defining \begin{equation}
\theta_{a}=\sum_{k=5}^{7}\mathcal{P}_{a}^{rs}\alpha_{k}(\mathbf{x},ct);\,\,\,\left(k,r,s\right)=\left(5,1,4\right),\left(6,2,4\right),\left(7,3,4\right)\label{eqn.23}\end{equation}
 etc, then (\ref{eqn.22a}) can be expressed as \begin{equation}
G_{ab}=\left(X_{a}\theta_{b}+X_{b}\theta_{a}\right).\label{eqn.24}\end{equation}

\subsection{The emergence of the massive vector field\label{sub:5.2}}

Since $G_{ab}$ is the most general field which can be formed out
of the invariant subspace $\mathcal{A}_{sy,3}$, and since the single
element of $\mathcal{B}_{sy,1}$ is orthogonal to every element of
$\mathcal{A}_{sy,3}$, then operating $\mathcal{B}_{sy,1}$ onto (\ref{eqn.24})
gives immediately, \[
X^{i}X^{j}G_{ij}\equiv\frac{\partial^{2}G^{ij}}{\partial x^{i}\partial x^{j}}=0,\]
 from which it immediately follows \begin{equation}
\frac{\partial G_{a}^{j}}{\partial x^{j}}=J_{a}~~\textrm{where}~~\frac{\partial J_{i}}{\partial x_{i}}=0,\label{eqn.24a}\end{equation}
 for some unspecified current $J$. Using (\ref{eqn.24}) this latter
equation can be expressed as \begin{equation}
\frac{\partial}{\partial x_{j}}\left(\frac{\partial\theta_{j}}{\partial x^{a}}+\frac{\partial\theta_{a}}{\partial x^{j}}\right)=J_{a}.\label{eqn.25}\end{equation}
 However, by (\ref{Intro2}), we have $X^{i}\mathcal{P}_{i}^{rs}\equiv0$
so that, from (\ref{eqn.23}), we get \begin{equation}
X^{j}\theta_{j}\equiv\frac{\partial\theta_{j}}{\partial x_{j}}=0,\label{eqn.26}\end{equation}
 so that (\ref{eqn.25}) becomes \begin{equation}
\Box\theta_{a}=-J_{a}.\label{eqn.g}\end{equation}
 However, since $\partial J_{i}/\partial x_{i}=0$ and $\partial\theta_{i}/\partial x_{i}=0$,
we can write $J_{a}=\kappa^{2}\theta_{a}+J_{a}^{0}$ for $\kappa^{2}\equiv m^{2}c^{2}/\hbar^{2}$
and conserved current $J_{a}^{0}$; finally, therefore, (\ref{eqn.g})
can be written as \begin{equation}
\Box\theta_{a}=-\frac{m^{2}c^{2}}{\hbar^{2}}\theta_{a}-J_{a}^{0}\label{eqn.g2}\end{equation}
 so that $\theta_{a}$, and hence $\mathcal{A}_{sy,3}$ also, represents
a massive vector field.

\subsection{The potential $\textbf{C}$ represents a massive vector field\label{subsec:5.3}}

The object \begin{equation}
\theta_{a}=\sum_{k=5}^{7}\mathcal{P}_{a}^{rs}\alpha_{k}(x);\,\,\,\left(k,r,s\right)=\left(5,1,4\right),\left(6,2,4\right),\left(7,3,4\right)\label{eqn.g4}\end{equation}
 has been identified as a non-zero mass four-vector field with three
degrees of freedom expressed in terms of the potentials $\textbf{C}\equiv\left(\alpha_{5},\alpha_{6},\alpha_{7}\right)$.
We show that $\textbf{C}$ also has the property of mass: (\ref{eqn.g4})
can be written explicitly as \[
\theta_{a}=\mathcal{P}_{a}^{14}\alpha_{5}+\mathcal{P}_{a}^{24}\alpha_{6}+\mathcal{P}_{a}^{34}\alpha_{7}\]
 which, after expanding the operators $\mathcal{P}_{a}^{rs}$ gives\begin{eqnarray}
(\theta_{1},\theta_{2},\theta_{3},\theta_{4}) & = & \left[-X_{4}(\alpha_{5},\alpha_{6},\alpha_{7}),\nabla\cdot(\alpha_{5},\alpha_{6},\alpha_{7})\right]\nonumber \\
 & \downarrow\nonumber \\
(\Theta,\theta_{4}) & = & \left[-X_{4}\textbf{C},\nabla\cdot\textbf{C}\right]\label{eqn.model1}\\
 & \downarrow\nonumber \\
\Box(\Theta,\theta_{4}) & = & (-X_{4}\Box\textbf{C},\nabla\cdot\Box\textbf{C}).\label{eqn.model2}\end{eqnarray}
 We now consider how $\textbf{C}\equiv\left(\alpha_{5},\alpha_{6},\alpha_{7}\right)$
must be constrained to ensure (\ref{eqn.model2}) assumes the form
of (\ref{eqn.g2}). A consideration of (\ref{eqn.model2}) shows that
the condition $\Box\textbf{C}=-\kappa^{2}\textbf{C}$, where $\kappa^{2}\equiv m^{2}c^{2}/\hbar^{2}$,
must hold since then, use of (\ref{eqn.model1}) reduces (\ref{eqn.model2})
to \[
\Box(\Theta,\theta_{4})=-\frac{m^{2}c^{2}}{\hbar^{2}}(\Theta,\theta_{4})\]
 which is (\ref{eqn.g2}) without the conserved current. That is,
each of the components of the potential $\textbf{C}\equiv\left(\alpha_{5},\alpha_{6},\alpha_{7}\right)$
must satisfy the Klein-Gordon equation so that $\textbf{C}$ is a
massive field.

\subsection{Transformation law}

In appendix \ref{subsec:Cov} we give an explicitly covariant representation
for the potentials $\textbf{C}$ which form, subsequently, allows
a derivation of the non-trivial transformation law for $\textbf{C}\rightarrow\hat{\textbf{C}}$
under $x\rightarrow\hat{x}=T(x-x_{0})$.

\section{Contact action in RMP electrodynamics\label{sec:6} }

Our demonstration that classical electromagnetism can be expressed
purely in terms of a relativistic magnetic potential has also revealed
a relativistic symmetry principle according to which the relativistic
magnetic potential is one half of a duality, the other half being
a potential with the property of mass. The electromagnetic field tensor
is represented as $F_{ab}$ whilst the field tensor associated with
the massive vector field we denote as $G_{ab}$.

None of this has any effect on the predictions and descriptions ordinarily
associated with the \emph{motive action} of the Lorentz force since,
within RMP electrodynamics, these remain identical to those arising
within the classical formalism. The introduction of $G_{ab}$ does,
however, create one difference of major significance: when it comes
to the discussion of \emph{force}, the conventional picture uses the
`Lorentz force law', $f_{a}^{(em)}=J^{i}F_{ai}$, which is more accurately
described as a bundled-up restatement of the \emph{definitions }of
the electric and magnetic fields in terms of their mechanical actions
on charged particles. In other words, conventional electrodynamics
provides a way of talking about mechanical forces in which the electromagnetic
field is really a proxy for these mechanical forces, but does not
naturally provide a \emph{force law }in the Newtonian sense of \emph{action
and reaction are equal and opposite }for electrodynamic phenomena.
In the absence of such a natural provision, the assumption is that
the reaction to a charge accelerating under the action of a field
$F_{ab}$ is felt (in a retarded sense) in the current which generates
$F_{ab}$ in the first place. 

However, the presence of the massive vector field within RMP electrodynamics
provides a natural way of generalizing Newton's Third Law to electrodynamic
phenonema so that the `Lorentz force law' is supplemented with a force
of reaction, arising from the massive vector field, to give a \emph{local
}balance of forces - the massive vector field is the source of \emph{contact
reaction} to the classical Lorentz force so that this latter becomes
reinterpreted as a \emph{local action contact force}. The details
follow.

\subsection{The interpretation of $G_{ab}$\label{sub:6.1}}

The fields $F_{ab}$ and $G_{ab}$ are irreducibly related though
a symmetry principle. Furthermore, the electromagnetic field is associated
with conserved current, $J_{em}^{a}\equiv eV^{a}/c$ say whilst, according
to \S\ref{sub:5.2}, $G_{ab}$ is also associated with a conserved
current, $J_{mass}^{a}$ say. It follows that, corresponding to the
conventional Lorentz force\begin{equation}
f_{a}^{(em)}=J_{em}^{i}F_{ai}\,,\label{eqn:FL0}\end{equation}
 there is also the related force\[
f_{a}^{(mass)}=J_{mass}^{i}G_{ai}\]
 arising from the action of the massive vector field on the conserved
current $J_{mass}^{a}$. It follows that there is a `total system
force' given by \begin{equation}
\mathcal{F}_{a}\equiv f_{a}^{(em)}+f_{a}^{(mass)}.\label{eqn:FL1}\end{equation}
 But the validity of (\ref{eqn:FL0}), insofar as it describes the
four-force exerted on a charged particle by a given electromagnetic
field, is very well established. So, the question now becomes: \emph{How
are we to interpret} $f_{a}^{(mass)}$ \emph{and hence} $\mathcal{F}_{a}$\emph{?}

An obvious possibility is that $f_{a}^{(mass)}$ acts to maintain
an overall balance of four-momentum at the instantaneous location
of the charged particle - that is, a generalized Newton's Third Law
applies so that (\ref{eqn:FL1}) becomes\begin{equation}
\mathcal{F}_{a}\equiv f_{a}^{(em)}+f_{a}^{(mass)}\equiv J_{em}^{i}F_{ai}+J_{mass}^{i}G_{ai}=0\label{eqn:FL2}\end{equation}
 which can be interpreted as a classical description of the instantaneous
conservation of four-momentum in a contact-action between a charged
particle and some kind of massive vector particle.

However, since there must \emph{always} be at least two charged particles
interacting, the foregoing lacks detail: thus, for exactness, suppose
that there are just two interacting charged particles, $P_{0}$ and
$P_{1}$ say, with associated four-currents $J_{0,em}^{a}$, $J_{0,mass}^{a}$
at $P_{0}$ and $J_{1,em}^{a}$, $J_{1,mass}^{a}$ at $P_{1}$. Each
of them will generate fields, $F_{ab}^{(0)}$, $G_{ab}^{(0)}$, $F_{ab}^{(1)}$
and $G_{ab}^{(1)}$ respectively, which are involved in the four-momentum
conserving interactions of the system. At each particle, a four-momentum
balance equation similar to (\ref{eqn:FL2}) and created from various
of these four component fields and currents must hold. Now, because
the $F_{ab}$-fields are massless whilst the $G_{ab}$-fields are
massive then, for example, $F_{ab}^{(0)}$ and $G_{ab}^{(0)}$ - both
generated by the currents at $P_{0}$ - \emph{cannot} act simultaneously
at $P_{1}$ so that, following this example through, the four-momentum
balance equation for $P_{1}$ cannot have the structure \[
\mathcal{F}_{a}^{(1)}\equiv J_{1,em}^{i}F_{ai}^{(0)}+J_{1,mass}^{i}G_{ai}^{(0)}=0.\]
 By contrast, $F_{ab}^{(0)}$ and $G_{ab}^{(1)}$ \emph{can} act simultaneously
at $P_{1}$ so that an obvious possibility for four-momentum balance
at $P_{1}$ is\begin{equation}
\mathcal{F}_{a}^{(1)}\equiv J_{1,em}^{i}F_{ai}^{(0)}+J_{1,mass}^{i}G_{ai}^{(1)}=0,\label{eqn:FL3}\end{equation}
 with the corresponding equation holding for four-momentum balance
at $P_{0}$. 

It is clear that, according to (\ref{eqn:FL3}), energy-momentum is
drawn from the massive vector field at $P_{1}$ and transferred directly
to the charged particle at $P_{1}$ (the current $J_{1,em}^{i}$).
There is therefore no requirement for energy-momentum transport between
$P_{0}$ and $P_{1}$ so that the role of the electromagnetic field,
$F_{ab}^{(0)}$, becomes that of a messenger carrying instructions
between $P_{0}$ and $P_{1}$ - that is, the role of the electromagnetic
field is strictly limited to its familiar, well understood, and verified
role \emph{of carrying information}. The energy required to enact
these instructions at $P_{1}$ has its source in the massive vector
field at $P_{1}$ and the linear momentum subsequently imparted to
the charged particle at $P_{1}$ is balanced in the reaction of the
massive vector field to the acceleration of this charged particle.
Thus, each of the charged particles \emph{causes} work to be done
on the other, but the source of the energy which does the work is
the massive vector field local to each of the charged particles -
in effect, we can think of the massive vector field as a material
{}``electromagnetic vacuum resource''.

To summarize, if the source of the energy-momentum which gives rise
to the acceleration of a charged particle is local to the particle
in the {}``electromagnetic vacuum resource'', then the electromagnetic
field itself has no role as a carrier of energy-momentum. Its function
can only be that of carrying coded {}``instructions'' between interacting
particles, telling each how much energy-momentum must be drawn from
the local environment.

\subsection{Implications for ideas of energy-momentum conservation within the
conventional formalism}

An immediate consequence of the foregoing is that the requirement
for the electromagnetic field to carry away energy-momentum to balance
the mechanical action of the Lorentz force disappears - that is, the
Maxwell stress tensor and Poynting's vector become formalisms representing
the propagation of information alone. To many physicists, used to
long practise of thinking in terms of the Maxwell stress tensor and
Poynting's vector as being `real' things, such a statement may appear
wild to the point of recklessness. To counter such appearances, it
is useful to recall the tautological nature of momentum-energy conservation
for the field-charge interaction as conceived within the conventional
formalism:

Begin by considering the energy-momentum tensor corresponding to the
flow of charge in an electromagnetic field in flat space written as\[
T_{n}^{m}\equiv S_{n}^{m}+\Theta_{n}^{m}\]
where $S_{n}^{m}$ is the electrodynamic component, defined in terms
of the electromagnetic field tensor, and $\Theta_{n}^{m}$ is the
mechanical component, defined in terms of the mechanical properties
of the charged particles comprising the current. Since the details
entail the result $T_{n,i}^{i}=0,$ then energy-momentum is said to
be conserved in the system. 

But now we recall the basic definition of the electric field at a
point as the \emph{mechanical force exerted on a unit charge placed
at the point} and the correspondingly similar definition of the magnetic
field in terms of its mechanical action on a unit charge. The Lorentz
`force law', $\mathcal{F}_{a}\equiv J^{i}F_{ai}$, is simply a bundled-up
version of these respective definitions of the electric and magnetic
fields - and is not\emph{ }a\emph{ law} of electromagnetic force.
Recall now that the $S_{n}^{m}$ component of $T_{n}^{m}$ is formed
by an algebraic process from $J^{i}F_{ai}\equiv\kappa F_{\,,j}^{ij}F_{ai}$
and that the mechanical component, $\Theta_{n}^{m}$, is formed by
an algebraic process from the mathematical representation of the observed
rates of change of the mechanical properties of the current charges.
It follows immediately that the identity $T_{n,i}^{i}=0$ is simply
a direct consequence of the fact that $S_{n}^{m}=-\Theta_{n}^{m}$
\emph{by definition}. That is, `energy-momentum conservation' as conventionally
defined for a system consisting of a flow of charge in an electromagnetic
field is a tautology and does not represent any basic law of physics. 

Finally, the idea that the $S_{n}^{m}$ component of this tautological
definition of $T_{n}^{m}$ derives from a free electromagnetic field
radiating real momentum and energy away from the location of the action
of the Lorentz force in response to this action is simply an additional
hypothesis justified \emph{only }by the absolute necessity to have
some mechanism of balancing momentum and energy.

\subsection{The possibility of scatter-free interactions}

The generalized Newton's Third Law of (\ref{eqn:FL3}), given by

\[
J_{1,em}^{i}F_{ai}^{(0)}+J_{1,mass}^{i}G_{ai}^{(1)}=0,\]
provides - in simple interactive systems at least - the possibility
of scatter-free redshifting of the electromagnetic field.

To see this, imagine a single electron at $\mathcal{P}_{1}$ having
a propagating electromagnetic field incident upon it. The Lorentz
force acts, in the ordinary way, to accelerate the electron, thereby
doing work upon it. According to the generalized Newton's Law above,
this energy must have its source in the massive vector field, represented
by $G_{ab}^{(1)}$. Similarly, the change in momentum experienced
by the electron must be balanced by the change in momentum experienced
by the massive vector particles associated with $G_{ab}^{(1)}$.

We interpret the detailed interaction as follows: 
\begin{enumerate}
\item work is done at $P_{0}$ to create the propagating electromagnetic
field $F_{ab}^{(0)}$; 
\item the energy expended in creating this field goes into the {}``vacuum''
and the amount of energy so expended - $E^{(0)}$ say - is recorded
as part of the information carried by $F_{ab}^{(0)}$;
\item the field $F_{ab}^{(0)}$is incident with the electron at $\mathcal{P}_{1}$
and the information that an amount of energy $E^{(0)}$ has been deposited
into the {}``vacuum'' puts an upper limit on the amount of work
that can be done on the electron where the energy required for this
work is drawn \emph{from} the {}``vacuum'' via the massive vector
field;
\item the electron is accelerated by the action of the Lorentz force in
a direction which is partly determined by the instantaneous velocity
of the electron at the moment of interaction - this is the stochastic
component which, \emph{in the classical picture}, gives rise to the
stochastic scattering of the incident electromagnetic wave required
for overall momentum conservation;
\item however, according to the generalized Newton's Law, overall momentum
conservation is guaranteed not by the scattering of electromagnetic
waves, but by the production of momentum in the massive vector field;
\item the outgoing wave need only contain a record of the energy balance
\emph{remaining }in the {}``vacuum'' after this interaction but
the momentum component, carrying information about the original source
of the wave, can remain unchanged - simply because the change in momentum
experienced by the electron can be balanced by the corresponding change
in momentum of the massive vector field \emph{and does not require
of necessity any change in the momentum information carried by the
wave. }In other words, the ongoing wave can remain unscattered by
the interaction.
\end{enumerate}

\section{Comments on free field quantization of the RMP formalism}

It was shown in \S\ref{sub4.3}, that the total field, $H_{ab}$,
of the RMP formalism lies in $\left\{ \mathcal{A}_{sk,3}\cup\mathcal{A}_{sy,3}\right\} $
so that\[
H_{ab}\equiv F_{ab}+G_{ab}\equiv\sum_{k=2}^{4}\mathcal{U}_{ab}^{k}\alpha_{k}(x)+\sum_{k=5}^{7}\mathcal{U}_{ab}^{k}\alpha_{k}(x),\,\,\,\,\mathcal{U}_{ab}^{k}\in\left\{ \mathcal{A}_{sk,3}\cup\mathcal{A}_{sy,3}\right\} \]
where the electromagnetic component, $F_{ab}$, is identical with
its canonical form in the temporal gauge. Since $F_{ab}$ and $G_{ab}$
are constrained only by the requirement that (\ref{eqn:FL3}) is satisfied
(four-momentum is locally conserved), and since there is no gauge
invariance associated with the massive $G_{ab}$ field, then an obvious
approach to quantizing $H_{ab}$ is to apply: 
\begin{itemize}
\item the canonical temporal gauge quantization procedure to $F_{ab}$;
\item the canonical quantization procedure for a massive vector field to
$G_{ab}$; 
\item a constraining condition between these two quantized fields, and the
currents with which they interact, equivalent to (\ref{eqn:FL3}),
$J_{1,em}^{i}F_{ai}^{(0)}+J_{1,mass}^{i}G_{ai}^{(1)}=0.$ 
\end{itemize}

\subsection{The $F_{ab}$ field}

The quantization conditions for $F_{ab}$ in the temporal gauge are
particularly straightforward: writing $\left(\alpha_{5},\alpha_{6},\alpha_{7}\right)\equiv\left(A_{1},A_{2},A_{3}\right)$
for the RMP and noting that the electric field is given by $\left(E_{1},E_{2},E_{3}\right)\equiv-\left(\dot{A}_{1},\dot{A}_{2},\dot{A}_{3}\right)$
then the equal time quantization conditions for $F_{ab}$ in the temporal
gauge are very simply expressed as\begin{eqnarray*}
\left[A_{i}(\mathbf{x},t),E_{j}(\mathbf{y},t)\right] & = & -i\delta_{ij}\delta^{3}(\mathbf{x}-\mathbf{y}),\\
\left[A_{i}(\mathbf{x},t),A_{j}(\mathbf{y},t)\right] & = & 0\\
\left[E_{i}(\mathbf{x},t),E_{j}(\mathbf{y},t)\right] & = & 0\end{eqnarray*}
which, in its elegant simplicity in using the three components of
the RMP directly, lends weight to the idea that the RMP formalism
is the fundamental expression of the classical em field.

\subsection{The $G_{ab}$ field}

The $G_{ab}$ field is shown in \S\ref{sec.vecboson} to be derived
from the three-component potential $\mathbf{C}\equiv\left(\alpha_{5},\alpha_{6},\alpha_{7}\right)$
and, in \S\ref{subsec:5.3}, we showed that each of the components
of this potential independently satisfied the Klein-Gordon equation.
Since there are no constraint between the three components of $\mathbf{C}$,
then we can immediately identify the free-particle canonical fields
as $C_{i}(\mathbf{x},t),\, i=1,2,3$ and conjugate fields as $\dot{C}_{i}(\mathbf{x},t),\, i=1,2,3$
so that we can immediately write down the equal time quantization
conditions for $G_{ab}$ as \begin{eqnarray*}
\left[C_{i}(\mathbf{x},t),\dot{C}_{j}(\mathbf{y},t)\right] & = & i\delta_{ij}\delta^{3}(\mathbf{x}-\mathbf{y}),\\
\left[C_{i}(\mathbf{x},t),C_{j}(\mathbf{y},t)\right] & = & 0\\
\left[\dot{C}_{i}(\mathbf{x},t),\dot{C}_{j}(\mathbf{y},t)\right] & = & 0\end{eqnarray*}
However, from (\ref{eqn.model1}), we have $\theta_{i}(\mathbf{x},t)=-\dot{C}_{i}(\mathbf{x},t),\, i=1,2,3$,
so that these conditions can be expressed as\begin{eqnarray*}
\left[C_{i}(\mathbf{x},t),\theta_{j}(\mathbf{y},t)\right] & = & -i\delta_{ij}\delta^{3}(\mathbf{x}-\mathbf{y}),\\
\left[C_{i}(\mathbf{x},t),C_{j}(\mathbf{y},t)\right] & = & 0\\
\left[\theta_{i}(\mathbf{x},t),\theta_{j}(\mathbf{y},t)\right] & = & 0\end{eqnarray*}

\subsection{The constraining condition and mass normalization}

The constraint \begin{equation}
J_{1,em}^{i}F_{ai}^{(0)}+J_{1,mass}^{i}G_{ai}^{(1)}=0.\label{eqn7.3.1}\end{equation}
 gives rise to an interesting situation: specifically, it implicitly
contains two characteristic masses, $M_{e}$, for the electron mass
and $M_{m}$ for the mass of the particle associated with the field
$G_{ab}^{(1)}$ and, in effect, provides a definition for the mass
ratio $M_{e}/M_{m}$. Thus, in any mass renormalization procedure,
$M_{e}\rightarrow M_{e}+\delta M_{e}$ and $M_{m}\rightarrow M_{m}+\delta M_{m}$,
we must have\[
\frac{M_{e}}{M_{m}}=\frac{M_{e}+\delta M_{e}}{M_{m}+\delta M_{m}}.\]
But, bearing in mind that measures of mass are always defined in terms
of the ratio of one mass to another, then the above appears to imply
that, so long as the masses appearing in each of the quantized $F_{ab}$
and $G_{ab}$ fields are normalized to appear in the ratio $M_{e}/M_{m}$
with a value defined according to (\ref{eqn7.3.1}), then any subsequent
mass renormalization procedure must be redundant.

\section{Electromagnetic fields and astrophysics}

Given that the force-law (\ref{eqn:FL3})\foreignlanguage{english}{
makes necessary a reinterpretation of }the Lorentz force as an instantaneous
local-action force - rather than as a retarded action-at-a-distance
force - then we can reasonably suspect it has the potential to provide
a new understanding of the loss mechanisms involved in any interaction
of an electromagnetic field with any system of charged particles.
Whilst the nature of the cosmological redshift is considered to be
fully understood within the context of classical Big-Bang cosmology
its origin has, in fact, been a topic of vigorous debate since its
discovery by Slipher in 1910 - the opposing view being, broadly speaking,
that cosmological redshift arises as a consequence of some form of
electrodynamical interaction with intervening material.

If there were no evidence whatsoever that a redshift effect can be
caused by anything other than universal expansion (or gravitational),
then it would be difficult to sustain a theoretical argument to the
contrary. However, there is \emph{at least }one redshift phenomenon
which is inarguably caused by some other mechanism - this is the $k$-effect
which manifests itself as an excess redshift in type $O$ and $B$
stars relative to other stars.

\subsection{The background radiation revisited}

The cosmic background radiation is widely thought of as a phenomenon
of Big-Bang cosmology and the prediction of its existence, before
the event of its detection and temperature measurement ($T\approx2.7^{o}K$),
is commonly associated with the name of Gamow \cite{key-13} who,
variously with Alpher and Herman \cite{key-5,key-6}, used the theory
to predict the putative temperatures of $T\approx5^{o}K$, $T\geq5^{o}K$,
$T\approx7^{o}K$ and $T\approx50^{o}K$ in the late 1940s and 1950s. 

However, against this, there is a much older narrative which has its
beginnings in the year of 1879 and Stefan's discovery of the law which
bears his name relating the temperature of a body in radiative equilibrium
to the energy flux density in which the body is immersed. At the time,
of course, it was commonly believed that the universe was stationary
and of infinite extent - and therefore in a state of dynamical equilibrium.
This latter belief, together with Stefan's Law, made it natural to
conceive the notion of \emph{the temperature of space }- by which
was meant the steady-state temperature of any suitably large passive
body placed in space and remote from any individual stars - and to
calculate its value from various points of view. 

The first recorded such calculation was made by Guilluame \cite{key-14}
in 1896; he, of course, had no notion of external galaxies and his
calculation was based purely on the presumed thermalization of our
own galaxy's starlight - even so, he obtained $5.6^{o}K$. The next
estimate of which we have knowledge was made by Eddington \cite{key-9}
in 1926 and - still presuming the thermalization of our own galaxy's
starlight, but having access to better data - he found $3.2^{o}K$.
By the 1930s, Regener's work \cite{key-19} on cosmic rays led him
to conclude that the energy density of the cosmic ray flux (reasoned
to be extragalactic by virtue of its isotropy) and that of the flux
of our own galaxy's starlight were essentially the same and, conseqently,
in equilibrium with each other. On this basis, he was able to use
his cosmic ray measurements to arrive at an estimate of $2.8^{o}K$
for the temperature of space. In a similar vein, in 1941 Herzberg
\cite{key-15} (based on observations made by McKellar) noted that,
supposing interstellar cyanogen to be immersed in an equilibrium heatbath,
the observed excitation of its molecules yielded a temperature of
$2.3^{o}K$ for that heatbath. 

So, we see that prior to the Gamow's predictions made on the basis
of the Big-Bang, the earlier concensus - centering on the idea of
the equilibrium universe - had already developed the concept of a
radiative equilibrium background and had, moreover, made extremely
accurate predictions of its temperature using a variety of quite different
physical systems. However, whilst it cannot be claimed that Gamow
and collaborators managed to predicted the background temperature
with any accuracy (unlike the proponents of the equilibrium universe),
it is also true that the equilibrium radiation background predicted
to exist by Gamow on the basis of the Big-Bang theory is \emph{in
addition to }the radiative fluxes (starlight, cosmic rays etc) used
by the earlier authors to determine the \emph{temperature of space}
so that, in this sense, Gamow did predict the existence of something
new. But, as we shall now point out, the Big-Bang theory is actually
superfluous even to this latter prediction.

\subsection{Cosmological redshift revisited}

We return to a very old problem, concerning the nature of the cosmological
redshift which, by definition, is a phenomenon affecting electromagnetic
radiation - and which is the primary concern of this paper. At the
time of its discovery, by Slipher \cite{key-25,key-26}, and the subsequent
recognition by Hubble \cite{key-27} that the degree of redshift observed
in the light emitted from a given object is a function of that object's
radial displacement from the observer, the only conceivable mechanism
to explain it was some form of the Doppler effect; but this interpretation
implied that the distant galaxies were all `flying apart', which was
in direct contradiction to the received view of the time.

A resolution of the conundrum was provided by Lema\^itre \cite{key-1}
who showed that a simple dust-model cosmology derived from Einstein's
General Relativity required that `spacetime' was expanding - and it
was this which provided what appeared to be a natural quasi-Doppler
explanation to the problem. It was Lema\^itre's model which led directly
to the idea of the hot Big-Bang and, subsequently, to Gamow's prediction
of an `afterglow' - the cosmic background radiation, discussed above. 

However, it is also obvious that \emph{within the context of an equilibrium
universe} the cosmological redshift is, of itself, prima-facie\emph{
}evidence for a loss of energy from the redshifted light and therefore
for the existence of an associated radiative flux which balances this
lost energy. The assumption of universal equilibrium - already strongly
supported by the earlier calculations discussed above - then requires
that this newly inferred radiative flux is also in equilibrium and
at a temperature of about $2.8^{o}K$. 

The only thing missing from such a narrative would have been the \emph{mechanism}
of redshifting. It is a matter of record that Hubble himself was never
convinced by the expanding universe explanation and, in the decades
following Lema\^itre's analyses, several authors suggested an alternative
generic mechanism, that of \textbf{\emph{tired light}}, by means of
which `travelling photons' were envisaged as interacting with other
particles, thereby losing energy and becoming redshifted. The first
such suggestion to appear in the literature (so far as we know) was
that of Nernst who, as early 1912, had developed the idea of a stationary
universe \cite{key-16}, subsequently suggesting that light was absorbed
by the \emph{luminiferous aether} \cite{key-17,key-18} to produce
the redshift effect. On this basis, he calculated a temperature for
space of about $1^{o}K$ - although it must be recognized that this
calculation depended to a great extent on assumptions made about the
nature of the proposed luminiferous aether. The post-war years saw
a renewed interest in the idea with, for example, Finlay-Freundlich
\cite{key-10,key-11,key-12} proposing a mechanism involving purely
photon-photon interactions and, on the basis of this, estimating the
temperature of space (in the equilibrium universe) to be in the range
$2^{o}K<T<6^{o}K$. Max Born \cite{key-7} subsequently pointed out
that the radiative flux associated with the Finlay-Freundlich mechanism
would necessarily be in the radio spectrum with $\lambda\approx15cm$;
in fact, we now know that the CBR has a characteristic wave length
of $\lambda\approx7cm$ so that Finlay-Freundlich's idea actually
had a real predictive success - not recognized at the time, of course.
However, the problem with the generic class of tired-light mechanisms
was simple: images of galaxies from which the light is redshifted
appear very sharp, whereas is seemed clear that any conceivable photon-particle
interactions must necessarily scatter the photons leading to the vastly
degraded image quality. Considerable effort was put into hypothesizing
various detailed mechanisms of this type, but none of those suggested
could stand up to critical analysis.

Now however, there is something entirely new to consider: specifically,
the present analysis allows for the interpretation of the Lorentz
force as a local action contact force, and it is this which presents
an opportunity for the reconsideration of loss mechanisms in electrodynamics.

\section{Conclusions}

The necessary inclusion of the massive vector field into electrodynamical
description leaves the Lorentz force unchanged in the sense that \emph{the
motive effects on currents of the Lorentz force in RMP electrodynamics
are identical to those of its effects in classical electrodynamics}
- thus, all of the successes of classical electrodynamics are captured
by RMP electrodynamics. But a fundamental difference is created -
specifically, the Lorentz force of RMP electrodynamics becomes explicitly
a \emph{local action contact force} with the massive vector field
playing the role of the `reaction particle'. At this point, given
that the photon as conceived in classical qed is the massless gauge
particle which arbitrates four-momentum conserving interactions between
charged particles, it becomes lucidly clear that the really significant
differences between the RMP formalism and the four-vector potential
formalism will arise in the domain of quantum electrodynamics - and
this is a study for the future.

\appendix
%dummy comment inserted by tex2lyx to ensure that this paragraph is not empty

\section{Reversibility: a further transformation property of the RMP \label{AppendixA}}

It is not immediately obvious that the transformation of the RMP described
above in \S\ref{sub2.3} is reversible as it should be. We show here,
by listing the steps implicit in $(\phi_{1},\phi_{2},\phi_{3})\rightarrow(\hat{\theta}_{1},\hat{\theta}_{2},\hat{\theta}_{3})\rightarrow(\phi_{1},\phi_{2},\phi_{3})$
, that it is.
\begin{enumerate}
\item \label{step1}Define $\Phi\equiv(\Phi_{1},\Phi_{2},\Phi_{3},\Phi_{4})=(\phi_{1},\phi_{2},\phi_{3},0)$; 
\item Rotation: $\hat{\Phi}_{r}=T_{r}^{i}\Phi_{i}$; 
\item Shift: $\hat{\theta}_{r}=\hat{\Phi}_{r}-\left(\hat{X}_{r}/\hat{X}_{4}\right)\hat{\Phi}_{4}$;
Note that $\hat{\theta}_{4}=0$ automatically; 
\item Inverse rotation: $\theta_{r}=T_{r}^{i}\hat{\theta}_{i}=T_{r}^{i}\hat{\Phi}_{i}-\left(T_{r}^{i}\hat{X}_{i}/\hat{X}_{4}\right)\hat{\Phi}_{4}=\Phi_{r}-\left(X_{r}/\hat{X}_{4}\right)\hat{\Phi}_{4}$; 
\item Shift: $\psi_{r}=\theta_{r}-\left(X_{r}/X_{4}\right)\theta_{4}=\left(\Phi_{r}-\left(X_{r}/\hat{X}_{4}\right)\hat{\Phi}_{4}\right)-\left(X_{r}/X_{4}\right)\left(\Phi_{4}-\left(X_{4}/\hat{X}_{4}\right)\hat{\Phi}_{4}\right)=\Phi_{r}$
since $\Phi_{4}=0$ by definition at step (\ref{step1}) above. 
\end{enumerate}
That is, $(\phi_{1},\phi_{2},\phi_{3})\rightarrow(\hat{\theta}_{1},\hat{\theta}_{2},\hat{\theta}_{3})\rightarrow(\phi_{1},\phi_{2},\phi_{3})$
as required.

\section{The Eigenspaces\label{Appendix:B}}

\subsection{Eigenspace $\mathcal{B}_{sy,1}$, $\lambda=2$}

\label{subsec.eigen1} $\mathcal{B}_{sy,1}$ is a one-dimensional
subspace of eigenvectors associated with the eigenvalue $\lambda=2$
and the subspace is defined by the single operator \begin{equation}
\mathcal{U}_{ab}^{1}\equiv X_{a}X_{b}\label{eqn.6}\end{equation}
 which is \textit{\emph{symmetric}} with respect to the indices $(a,b)$.

\subsection{Eigensubspace $\mathcal{A}_{sk,3}$, $\lambda=1$}

\label{subsec.eigen4} $\mathcal{A}_{sk,3}$ is a three-dimensional
subspace of skew-symmetric eigenvectors associated with the eigenvalue
$\lambda=1$ and a basis for the subspace is given by \begin{equation}
\mathcal{U}_{ab}^{k}=\left(X_{a}\delta_{b}^{r}-X_{b}\delta_{a}^{r}\right),~~~k=2,3,4\label{eqn.9}\end{equation}
 where, for $k=(2,3,4)$ then $r$ takes any three distinct values
from the set $(1,2,3,4)$; for example, $r=(1,2,3)$; these eigenvectors
are \textit{\emph{skew-symmetric}} with respect to the indices $(a,b)$.

\subsection{Eigensubspace $\mathcal{A}_{sy,3}$, $\lambda=1$}

\label{subsec.eigen5} $\mathcal{A}_{sy,3}$ is a three-dimensional
subspace of symmetric eigenvectors associated with the eigenvalue
$\lambda=1$ and a basis for the subspace is given by

\begin{eqnarray}
\mathcal{U}_{ab}^{k} & = & X_{a}\left(X^{r}\delta_{b}^{s}-X^{s}\delta_{b}^{r}\right)+X_{b}\left(X^{r}\delta_{a}^{s}-X^{s}\delta_{a}^{r}\right),~~~k=5,6,7\label{eqn.10}\end{eqnarray}
 where for $k=(5,6,7)$, then $(r,s)$ is three distinct pairs chosen
from $(1,2,3,4)$. The basis is most conveniently chosen by picking
any one of the four digits and pairing it with the remaining three:
for example, $(r,s)=(1,4),(2,4),(3,4)$. These eigenvectors are \textit{\emph{symmetric}}
with respect to the indices $(a,b)$.

\subsection{Eigensubspace $\mathcal{C}_{sk,3}$, $\lambda=0$}

\label{subsec.eigen3} $\mathcal{C}_{sk,3}$ is a three-dimensional
subspace of skew-symmetric eigenvectors associated with the eigenvalue
$\lambda=0$ and a basis for the subspace is given by \begin{eqnarray}
\mathcal{U}_{ab}^{k} & = & \frac{X^{r}X^{s}X^{t}}{2\, X^{a}X^{b}}\left(\left(\delta_{a}^{r}-\delta_{a}^{s}\right)\left(\delta_{b}^{s}-\delta_{b}^{t}\right)-\left(\delta_{b}^{r}-\delta_{b}^{s}\right)\left(\delta_{a}^{s}-\delta_{a}^{t}\right)\right)\equiv\mathcal{D}_{ab}^{rst},\label{eqn.8}\\
k & = & 8,9,10\nonumber \end{eqnarray}
 where the numerical factor is inserted for convenience and where
typically, for $k=(8,9,10)$ then $(r,s,t)=(2,3,4),(1,3,4),(1,2,4)$;
these eigenvectors are \textit{\emph{skew-symmetric}} with respect
to the indices $(a,b)$.

\subsection{Eigensubspace $\mathcal{C}_{sy,6}$, $\lambda=0$\label{sub:B5}}

\label{subsec.eigen2} $\mathcal{C}_{sy,6}$ is a six-dimensional
subspace of symmetric eigenvectors associated with the eigenvalue
$\lambda=0$ and a basis for the subspace is given by \begin{equation}
\mathcal{U}_{ab}^{k}=\left(X^{r}\delta_{a}^{s}-X^{s}\delta_{a}^{r}\right)\left(X^{r}\delta_{b}^{s}-X^{s}\delta_{b}^{r}\right);\,\,\, k=11...16\label{eqn.7}\end{equation}
 where, typically, for $k=11...16$ then $(r,s)=(1,2),(1,3)(1,4),(2,3),(2,4),(3,4)$
and $\delta_{ab}$ is the $4\times4$ unit matrix; these eigenvectors
are \textit{\emph{symmetric}} with respect to the indices $(a,b)$.
As we have already noted, $\mathcal{C}_{sy,6}$ is the invariant subspace
of $\mathcal{S}_{16}$ which does not play any obvious part in the
electromagnetic theory being discussed here.

\section{An explicitly covariant formulation for the potentials $\left(\alpha_{5},\alpha_{6},\alpha_{7}\right)$\label{subsec:Cov}}

We have shown, in \S\ref{sub2.3}, the relationship between the RMP
and the explicitly covariant four-vector potential of the classical
formalism. We now perform a similar analysis to obtain an explicitly
covariant expression for the potentials $\left(\alpha_{5},\alpha_{6},\alpha_{7}\right)$
associated with $\mathcal{A}_{sy,3}$.

\subsection{Preliminaries}

From (\ref{eqn.22a}), (\ref{eqn.23}) and (\ref{eqn.24}) we have\begin{eqnarray}
V_{ab} & = & \sum_{k=5}^{7}(X_{a}\mathcal{P}_{b}^{rs}+X_{b}\mathcal{P}_{a}^{rs})\alpha_{k}(x)\equiv X_{a}V_{b}+X_{b}V_{a},\label{eqn29}\\
 &  & \left(k,r,s\right)=\left(5,1,4\right),\left(6,2,4\right),\left(7,3,4\right)\nonumber \end{eqnarray}
 We also have the notation $\left(C_{1},C_{2},C_{3}\right)\equiv(\alpha_{5},\alpha_{6},\alpha_{7})$
so that (\ref{eqn29}) can be written as\begin{equation}
V_{ab}=\sum_{r=1}^{3}\left(X_{a}\mathcal{P}_{b}^{r4}+X_{b}\mathcal{P}_{a}^{r4}\right)C_{r}\equiv\sum_{r=1}^{3}\mathcal{Q}_{ab}^{r4}C_{r}\,\,\,{\rm say}.\label{eqn29a}\end{equation}
 Although $V_{ab}$ is a covariant object, the objects from which
it is composed in (\ref{eqn29a}), $\mathcal{Q}_{ab}^{r4}$ and $C_{r},\, r=1..3$,
are not obviously so. To obtain a manifestly covariant expression
for $V_{ab}$, we begin by noting two significant properties of the
operators $\mathcal{Q}_{ab}^{rs}$:
\begin{itemize}
\item the elements of the set $\mathcal{Q}\equiv\left(\mathcal{Q}_{ab}^{rs},\,\, r=1..4,\, s=1..4\right)$
are easily shown to satisfy\begin{equation}
X^{r}\mathcal{Q}_{ab}^{st}+X^{t}\mathcal{Q}_{ab}^{rs}+X^{s}\mathcal{Q}_{ab}^{tr}\equiv0\label{eqn29f}\end{equation}
 where $(r,s,t)$ is any triple (no repetition) chosen from $(1,2,3,4)$
- for example, $(r,s,t)=(1,2,4).$ This implies, for example, that
$\left(\mathcal{Q}_{ab}^{12},\mathcal{Q}_{ab}^{14},\mathcal{Q}_{ab}^{24}\right)$
is a linearly \emph{dependent} set, but that $\left(\mathcal{Q}_{ab}^{14},\mathcal{Q}_{ab}^{24},\mathcal{Q}_{ab}^{34}\right)$
is not. It is now easily seen, by reference to (\ref{eqn29f}), that
this latter triple spans the space of $\mathcal{Q}$ - it represents
one possible basis of $\mathcal{Q}$. 
\item Since $\mathcal{P}_{a}^{rs}\equiv X^{r}\delta_{a}^{s}-X^{s}\delta_{a}^{r}$
transforms as a third-order tensor, then $\mathcal{Q}_{ab}^{rs}\equiv\left(X_{a}\mathcal{P}_{b}^{rs}+X_{b}\mathcal{P}_{a}^{rs}\right)$
transforms as a fourth-order tensor. 
\end{itemize}

\subsection{An explicitly covariant formulation for $V_{ab}$}

It is now obvious that \emph{if} we can find a mapping of the potentials
$(C_{1},C_{2},C_{3})\rightarrow\Phi_{ab}$, where $\Phi_{ab}$ is
a second-order tensor object such that $V_{ab}=\mathcal{Q}_{ab}^{ij}\Phi_{ij}$,
then the fact that $\left(\mathcal{Q}_{ab}^{14},\mathcal{Q}_{ab}^{24},\mathcal{Q}_{ab}^{34}\right)$
spans the space $\mathcal{Q}$ guarantees that the explicitly covariant
expression $\mathcal{Q}_{ab}^{ij}\Phi_{ij}$ can be expressed in terms
of $\left(\mathcal{Q}_{ab}^{14},\mathcal{Q}_{ab}^{24},\mathcal{Q}_{ab}^{34}\right)$,
so that the formal structure of (\ref{eqn29a}) would be preserved
under $x\rightarrow\hat{x}=T\left(x-x_{0}\right)$

We will suppose that such a $\Phi_{ab}$ exists: to deduce its structure,
we begin by noting that, since $\mathcal{P}_{a}^{rs}\equiv X^{r}\delta_{a}^{s}-X^{s}\delta_{a}^{r}$,
then $\mathcal{P}_{a}^{ij}\Phi_{ij}\equiv0$ for any \emph{symmetric
$\Phi_{ab}$.} It follows that $\Phi_{ab}$ must be skew-symmetric.
Furthermore, since we require that it results from a mapping $(C_{1},C_{2},C_{3})\rightarrow\Phi_{ab}$,
and therefore contains only three degrees of freedom, then it must
have the following invariant structure\begin{equation}
\Phi_{ab}=\left(\begin{array}{cccc}
0 & D_{1} & D_{2} & D_{3}\\
-D_{1} & 0 & D_{3} & D_{2}\\
-D_{2} & -D_{3} & 0 & D_{1}\\
-D_{3} & -D_{2} & -D_{1} & 0\end{array}\right),\label{eqn30}\end{equation}
 where $\left(D_{1},D_{2},D_{3}\right)$ are undetermined functions
of $\left(C_{1},C_{2},C_{3}\right)$. By (\ref{eqn.24}), this expression
for $\Phi_{ab}$ must satisfy \begin{equation}
V_{ab}=\left(X_{a}\mathcal{P}_{b}^{ij}+X_{b}\mathcal{P}_{a}^{ij}\right)\Phi_{ij}\equiv\left(X_{a}V_{b}+X_{b}V_{a}\right)\label{eqn29b}\end{equation}
 where $V_{a}\equiv\mathcal{P}_{a}^{ij}\Phi_{ij}$. We now have two
ways of expressing $V_{a}$: from (\ref{eqn.g4}), we have\begin{equation}
V_{a}=\sum_{r=1}^{3}\mathcal{P}_{a}^{r4}C_{r}\label{eqn29c}\end{equation}
 whilst from (\ref{eqn29b}), we have\begin{equation}
V_{a}=\mathcal{P}_{a}^{ij}\Phi_{ij}.\label{eqn29d}\end{equation}
 If these are to be equivalent, then the right-side of (\ref{eqn29c})
must equal the right-side of (\ref{eqn29d}). Thus, we get the requirement\begin{equation}
\left\{ V_{a}\right\} =\left(\begin{array}{c}
-X_{4}C_{1}\\
-X_{4}C_{2}\\
-X_{4}C_{3}\\
X_{1}C_{1}+X_{2}C_{2}+X_{3}C_{3}\end{array}\right)=\left(\begin{array}{c}
-X_{2}D_{1}-X_{3}D_{2}-X_{4}D_{3}\\
X_{1}D_{1}-X_{3}D_{3}-X_{4}D_{2}\\
X_{1}D_{2}+X_{2}D_{3}-X_{4}D_{1}\\
X_{1}D_{3}+X_{2}D_{2}+X_{3}D_{1}\end{array}\right).\label{eqn30a}\end{equation}
 Therefore, these two distinct expressions for $V_{a}$ are mutually
consistent if the solution of any three of the four component equations
of (\ref{eqn30a}) satisfies the fourth. The first three of these
equations give, after some algebra:\[
-X_{4}\left(X_{1}C_{1}+X_{2}C_{2}+X_{3}C_{3}\right)=-X_{4}\left(X_{1}D_{3}+X_{2}D_{2}+X_{3}D_{1}\right)\]
 which we see immediately is consistent with the fourth equation of
(\ref{eqn30a}). In other words, $\Phi_{ab}$ defined at (\ref{eqn30})
provides an explicitly covariant representation for the potentials
$\left(C_{1},C_{2},C_{3}\right)$ as required.

\subsection{Transformation properties of the potentials $\left(C_{1},C_{2},C_{3}\right)$}

We can now infer the transformation law for the potentials $\left(C_{1},C_{2},C_{3}\right)\rightarrow\left(\hat{C}_{1},\hat{C}_{2},\hat{C}_{3}\right)$
under $x\rightarrow\hat{x}=T\left(x-x_{0}\right)$:
\begin{enumerate}
\item $\left(C_{1},C_{2},C_{3}\right)\rightarrow\left(D_{1},D_{2},D_{3}\right)$
by solving the equations (\ref{eqn30a}); 
\item hence $\left(D_{1},D_{2},D_{3}\right)\rightarrow\Phi_{ab}$ from (\ref{eqn30}); 
\item $\Phi_{ab}\rightarrow\hat{\Phi}_{ab}$; 
\item $\hat{\Phi}_{ab}\rightarrow\left(\hat{D}_{1},\hat{D}_{2},\hat{D}_{3}\right)$
and hence $\left(\hat{D}_{1},\hat{D}_{2},\hat{D}_{3}\right)\rightarrow\left(\hat{C}_{1},\hat{C}_{2},\hat{C}_{3}\right)$
by solving (\ref{eqn30a}) again. 
\end{enumerate}

\section{The Dual Field of Electrodynamics From $\mathcal{C}_{sk,3}$\label{App:The-Dual-Field}}

The most general tensor generated by $\mathcal{C}_{sk,3}$ is given
by \begin{eqnarray}
G_{ab} & = & \sum_{8}^{10}\mathcal{U}_{ab}^{k}\alpha_{k}(x)\label{eqn.new1}\\
\mathcal{U}_{ab}^{k} & = & \frac{X^{r}X^{s}X^{t}}{X^{a}X^{b}}\left(\left(\delta_{a}^{r}-\delta_{a}^{s}\right)\left(\delta_{b}^{s}-\delta_{b}^{t}\right)-\left(\delta_{b}^{r}-\delta_{b}^{s}\right)\left(\delta_{a}^{s}-\delta_{a}^{t}\right)\right);\,\,\, k=8,9,10\nonumber \end{eqnarray}
 where, typically, for $k=(8,9,10)$ then $(r,s,t)=(2,3,4),(1,3,4),(1,2,4)$.
If, for the sake of convenience, we define $(\alpha_{8},\alpha_{9},\alpha_{10})\equiv(A_{1},A_{2},A_{3})$,
and use the given basis for $(r,s,t)$, then it is easily found that
\[
G_{ab}=\left(\begin{array}{cccc}
0 & -X_{4}A_{3} & X_{4}A_{2} & X_{2}A_{3}-X_{3}A_{2}\\
X_{4}A_{3} & 0 & -X_{4}A_{1} & X_{3}A_{1}-X_{1}A_{3}\\
-X_{4}A_{2} & X_{4}A_{1} & 0 & X_{1}A_{2}-X_{2}A_{1}\\
X_{3}A_{2}-X_{2}A_{3} & X_{1}A_{3}-X_{1}A_{2} & X_{2}A_{1}-X_{1}A_{2} & 0\end{array}\right)\]
 A consideration of this skew-symmetric object soon shows that it
is no more than a re-ordering of the terms of the electromagnetic
field tensor - which suggests an electrodynamic interpretation of
$G_{ab}$. In fact, it is easily shown that $G_{ab}=\epsilon_{abmn}F_{mn}$
where $\epsilon_{abmn}$ is the Levi-Civita permutation tensor. Thus,
$G_{ab}$ is the dual of $F_{ab}$.

\end{document}